\documentclass[epj]{svjour}
\usepackage{graphics,color}

\begin{document}

\title{Clusterized nuclear matter in the (proto-)neutron star crust 
and the symmetry energy}
\author{Ad. R. Raduta\inst{1}
\and
F. Aymard\inst{2,3}
\and 
F. Gulminelli\inst{2,3}
}                     
\institute{IFIN-HH, Bucharest-Magurele, POB-MG6, Romania
\and
CNRS, UMR6534, LPC ,F-14050 Caen c\'edex, France 
\and 
ENSICAEN, UMR6534, LPC ,F-14050 Caen c\'edex, France}
\date{Received: date / Revised version: date}
%
\abstract{
Though generally agreed that the symmetry energy plays a dramatic role
in determining the structure of neutron stars and the evolution of
core-collapsing supernovae, little is known in what concerns its value
away from normal nuclear matter density and, even more important,
the correct definition of this quantity in the case of unhomogeneous matter.
Indeed, nuclear matter traditionally addressed by mean-field 
models is uniform while clusters are known to exist in the dilute 
baryonic matter which constitutes 
the main component of compact objects outer shells.
In the present work we investigate
the meaning of  symmetry energy in the case of clusterized systems and
the sensitivity of the proto-neutron star 
composition and equation of state to the effective interaction. 
To this aim an improved Nuclear Statistical Equilibrium (NSE) model 
is developed, where the same 
effective interaction is consistently used to determine the clusters and 
unbound particles energy functionals in the self-consistent mean-field 
approximation. In the same framework, in-medium modifications to the 
cluster energies due to the presence of the nuclear gas are evaluated.
We show that the excluded volume effect does not exhaust the in-medium effects
and an extra isospin and density dependent energy shift has to be considered
to consistently determine the composition of subsaturation stellar matter.
The symmetry energy of diluted matter is seen to depend on the isovector
properties of the effective interaction, but its behavior with density and its 
quantitative value are strongly modified by clusterization.
} 

\maketitle

\section{Introduction}
\label{intro}

While the properties of the energy density as a function of the baryonic 
density $\rho$ and the isospin asymmetry 
$\delta=(\rho_n-\rho_p)/\rho$, {\it i.e.} the equation of state,  
close to saturation are constrained by the experimental data of atomic nuclei, 
little is known about their behavior away from the saturation density of 
symmetric nuclear matter $\rho_0^0$. 
This is especially true in the isovector sector. As a consequence, 
the so-called symmetry energy per baryon, 
defined as the curvature of the energy 
per baryon in the isospin direction calculated for symmetric matter, 
$2e_{sym}\equiv\partial^2 e/
\partial \delta^2(\rho,\delta=0)$,
is presently the object of intense research. 

This quantity is thought to impact on a variety
of phenomena ranging from nuclear masses \cite{moller}, 
neutron skin thickness \cite{prex},
fragment and particle production and flows in intermediate and high 
energy heavy-ion collisions 
\cite{colonna_physrep2005,baoan,russotto}, 
collective modes
\cite{pygmy,roca-maza_prc2013,liang_prc2007,trippa_prc2008} ,
structure and properties of neutron star crust 
\cite{steiner_2005,fattoyev_prc2012,grill_2012,gandolfi_prc2012}, 
just to cite a few. For recent reviews of these different topics,
see the corresponding articles of this volume.
Within the present constraints, still a broad range of behaviors
is put forward by the different effective interactions 
which can give consistent predictions for symmetric matter 
and still diverge in their isovector behavior, 
effectively measured by the slope
$L \equiv 3 \rho_0^0 \left( de_{sym}/d\rho\right)_{\rho=\rho_0^0}$
and curvature
$K_{sym} \equiv 9 \rho_0^2 \left( d^2 e_{sym}/d \rho^2\right)_{\rho=\rho_0^0}$
of the symmetry energy  around $\rho_0^0$.

For the idealized uncharged uniform system that nuclear matter (NM) 
is considered as, it is customary \cite{lopez_npa1988,bao-an_npa2001} 
to assume, at least close to symmetry $|\delta|\ll1$, a parabolic
dependence of the energy per baryon on the asymmetry parameter  as
$e(\rho,\delta) \approx e_0(\rho)+e_{sym}(\rho) \delta^2$.
Within this approximation, the symmetry energy gets the intuitive 
physical meaning of representing the energetic cost of converting 
isospin symmetric matter into neutron matter.

The validity of this representation stems from the charge invariance property 
of the strong interaction: in the absence of electromagnetic couplings 
symmetric matter ($\delta=0$) minimizes the energy  at any baryonic density 
$\rho$.
In turn, the absence of Coulomb effects in baryonic matter is due to the 
assumption of homogeneity associated to the thermodynamic limit.
 
Because the contrasting effects of Coulomb, which is minimized when 
matter is clusterized, and surface  energy, which is minimized in 
uniform matter, it is however very well known that baryonic matter, 
as it can be found in core-collapse supernovae explosions (CCSN) and 
(proto-)neutron stars (P)NS, possesses an inhomogeneous structure 
\cite{lattimer_2004,haensel_2007}. 
It basically consists  of a dense component (clusters), whose density 
is roughly the normal nuclear matter density, and a dilute component, 
constituted of unbound nucleons.
The fact that this does not correspond to the liquid-gas (LG) phase
coexistence that occurs in uncharged NM 
\cite{glendenning_2001,camille_npa2006},
but to a mixture where the 
two components alternate on a microscopic scale, is due to the electron 
screening, and makes the crust-core transition  a continuous one 
\cite{horowitz_prc2004,ens-ineq}. 

Because of this inhomogeneity, matter is locally charged, 
meaning that the energy density minimum might not be located at $\delta=0$,
and the parabolic expansion around symmetric matter might  not be justified.
Therefore, the meaning itself of symmetry energy is questionable.
Moreover, given that an important  fraction of matter is at (or close to) 
saturation nuclear density, where all realistic nucleon-nucleon interaction 
potentials provide for physical observables values in agreement with the 
experimental data, it is expected that the sensitivity of the symmetry 
energy to the underlying effective interaction might be partially or totally 
washed out.

The aim of the present work is to investigate the validity of the 
parabolic approximation, the meaning of the 
symmetry energy and the sensitivity to the EOS in the case of net-charge neutral
inhomogeneous nuclear matter at sub-saturation densities treated within the
nuclear statistical equilibrium (NSE) approach 
\cite{phillips_1994,souza_2009,botvina_npa2010,hempel_npa2010,noi_2010,blinnikov_2011}.

\section{Clusters in stellar matter}
\label{Custers_in_stars}

Self-consistent mean-field approaches have shown already thirty years ago 
that the clusterized  structure  typical of the outer crust of 
neutron stars persists at any density below saturation, with a continuous 
variation of cluster sizes, isospin and shape 
as the volume fraction of the dense phase increases 
\cite{negele_vautherin,williams_1985}. 
Confirmation is offered by microscopic calculations
\cite{maruyama_1998,horowitz_prc2004,sebille_prc2011} 
which additionally show that
the same stands true at finite temperature and for arbitrary proton fractions 
\cite{sonoda_2008,watanabe_2009,newton_prc2009}. 
Knowledge of both thermodynamical response of baryonic matter and its chemical
composition represents a chief requirement for astrophysical simulations of
CCSN and PNS cooling.
The task is challenging, as a wide range of temperatures
($10^9 <T<2 \cdot 10^{11}$ K), baryonic 
densities ($10^{5} < \rho < 10^{14}$ g/cm$^3$)
and proton fraction ($0 \leq Y_p \leq 1$) are spanned, with matter 
presenting a wealth of phenomenologies. 
Though in principle 
preferable, microscopic calculations are too expensive
to be exploited for such a task. 
It is considered that an acceptable compromise is offered by NSE-models
which describe matter as a mixture of loosely interacting nucleons 
and nuclei in thermal and chemical equilibrium 
\cite{phillips_1994,souza_2009,botvina_npa2010,hempel_npa2010,noi_2010,blinnikov_2011}. 
The basic idea behind NSE is the Fisher conjecture
that strong interactions in dilute matter may be completely exhausted by 
clusterization \cite{fisher}.

Several such models have been proposed in the last years. 
Allowing for a distribution of nuclear species, they represent a step 
forward with respect to the pioneering work of Lattimer and Swesty 
\cite{LS_npa91}, where only a unique representative nucleus had been 
considered. They are valid on the huge ranges of densities, 
temperature and proton fractions explored during the core collapsing 
supernovae and are, in principle, able to describe the inhomogeneous (crust)-
homogeneous (core) matter transition. 

A shortcoming of NSE models is the inconsistency among the energy 
functionals adopted for the description of unbound nucleons and nuclear species.
This is the case of our previous work \cite{noi_2010,ens-ineq} where we used 
the self-consistent mean-field treatment for unbound nucleons and a 
phenomenological liquid-drop parametrization for the cluster functional.
The same is true for the work of Ref. \cite{hempel_npa2010}, 
where a table of experimental binding energy was employed. 
The use of experimental binding energies should give in 
principle an optimal predictive power to the model, but most of the clusters
present in stellar matter lie beyond the neutron drip-line or the fission 
instability line in terrestrial laboratories. 
In such a situation a mass table has to be complemented by a theoretical 
prediction, and the problem of consistency with the treatment of 
continuum states arises again.

To describe clusterized baryonic matter at sub-saturation densities, 
we shall adopt in this work the non-relativistic density functional 
approach with Skyrme effective interactions.
The nuclear gas is then described by the (free) energy density obtained 
with these interactions in the homogeneous limit \cite{noi_2010}, 
while the cluster functional parameters are calculated from the 
same effective interaction as parametrized in Ref. \cite{danielewicz_npa818}.

Another limitation of NSE models is the phenomenological treatment
of hard-core interactions among the clusterized and gas components via
the excluded volume approximation. A recent comparison \cite{hempel-typel} 
of the excluded volume approach with a more sophisticated calculation of 
the binding energy shift due to the Pauli blocking effect of the continuum 
states shows that the semi-classical excluded-volume
gives a reasonably good overall description of the in-medium modifications, 
in particular correctly leading to the Mott
dissolution of clusters in a dense gas \cite{roepke}. 
We will argue in the next section that indeed,  within the 
local density approximation, the excluded volume effect completely accounts 
for the bulk part of the in-medium modification.

In-medium surface effects are completely neglected by the excluded volume 
approach.
Such effects can however be readily implemented if the cluster and gas 
functionals are derived from the same effective interaction. 
This extra correction is addressed in Section \ref{section:deltaEsurf}.

\subsection{The model}
\label{section:model}

At sub-saturation densities matter that constitutes CCSN and (P)NS is composed
of neutrons, protons, light and heavy bound clusters of nucleons, 
a charge neutralizing background of electrons and positrons 
(in pair-equilibrium) and photons in thermal and chemical equilibrium.
Depending on the thermodynamical conditions, neutrinos and anti-neutrinos
can also participate to the equilibrium.

As there is no interaction among leptons, photons and baryons other than
the electromagnetic one, the different systems are treated separately 
and their contributions  to the global thermodynamical potentials summed up.
Considering an arbitrary statistical ensemble and letting $F$ 
be its thermodynamical potential one writes,
\begin{equation}
f_{tot}=f_{baryon}+f_{lepton}+f_{\gamma},
\end{equation}
where we have replaced the extensive thermodynamical potential by its
density $f=\lim_{V \rightarrow \infty} F/V$. $V$ stands for the volume which, 
at the thermodynamical limit, is irrelevant.

In what regards the non-baryonic sectors, we shall adopt the 
traditional description valid in this $T-\rho$ range \cite{LS_npa91}:
leptons are considered to form an ideal highly relativistic gas 
in pair-equilibrium and photons are considered as an ultra-relativistic
Bose gas.

It is worthwhile to remind that the electron-chemical potential is imposed
by the net charge neutrality constrain $\rho_e=\rho_p$ through the relation
\begin{equation}
\rho_e=\frac{g_e}{6 \pi^2} \left( \frac{\mu_e}{\hbar c} \right)^3
\left[ 
1+ \frac1{\mu_e^2} \left( \pi^2 T^2 -\frac32 m_e^2 c^4\right)
\right],
\label{eq:mu_e}
\end{equation}
where $g_e$ and $m_e$ respectively stand for the electron spin degeneracy 
and rest mass.

In the specific application studied in this paper, we will consider
low temperature matter in $\beta$-equilibrium, appropriate for the description
of the (P)NS crust and the pre-bounce CCSN dynamics. 
The emphasis on low temperature will allow us to concentrate on the 
influence of the energy functional by minimizing the entropic contributions. 
In this physical situation, matter is completely transparent to neutrinos 
which do not participate to the chemical equilibrium.
This latter is then defined by the relation:
\begin{equation}
\mu_n+m_nc^2=\mu_p +m_p c^2 +\mu_e +m_e c^2.
\end{equation}

The baryonic sector is composed of various loosely interacting
nuclear species including unbound nucleons. The relative amount of 
clusterized and unbound components evolves continuously as a function of 
baryonic density, and the limiting structures correspond to a crystal 
(at low densities) and homogeneous matter of interacting nucleons 
(at $\rho \preceq  \rho_0^0$). 
The thermal and chemical equilibrium among the different nuclear species 
determines, together with mass and charge conservation and excluded volume 
constraints, the multiplicity of the different particles and clusters. 
In the present work we shall adopt the analytically tractable model proposed
in Ref. \cite{ens-ineq}.
In this model the gas of nucleons is treated in the bulk uniform limit 
within the mean-field approximation with Skyrme effective  interactions. 
The non-relativistic character allows to express the energy density in terms of
nucleon-nucleon coupling constants and single-particle species 
($\rho_g=\rho_{gn}+\rho_{gp}$, $\rho_{g3}=\rho_{gn}-\rho_{gp}$) 
and kinetic energy  ($\tau_g=\tau_{gn}+\tau_{gp}$, 
$\tau_{g3}=\tau_{gn}-\tau_{gp}$)  densities:
\begin{eqnarray}
\epsilon=\frac{\hbar ^{2}}{2m}\tau_g + C_{0}\rho_g ^{2}+D_{0}\rho _{g3}^{2}
+C_{3}\rho_g ^{\sigma +2}+D_{3}\rho_g ^{\sigma }\rho _{g3}^{2} \nonumber \\
+C_{eff}\rho_g \tau_g +D_{eff}\rho _{g3}\tau _{g3}
\label{eq:egas}
\end{eqnarray}
where the coefficients $C_{i}$ and $D_{i}$, associated respectively 
with the symmetry and asymmetry contributions, are linear combinations of 
the traditional Skyrme parameters, and
the unbound particle densities are given by: 
\begin{equation}
\rho_{gi}=\frac{4 \pi }{h^3} \left(\frac{2m^*_i}{\beta} \right)^{\frac 32} 
F_{\frac 12}(\beta \tilde \mu_i),
\end{equation}
 and 
\begin{equation}
\tau_{gi}=\frac{8\pi^3 }{h^5} \left( \frac{2 m^*_i}{\beta } \right )^{\frac 52}
F_{\frac 32}(\beta \tilde \mu_i),
\end{equation}
 with 
$F_{\nu}(\eta)=\int_0^{\infty} dx \frac{x^{\nu}}{1+\exp \left(x-\eta\right)}$ 
standing for the Fermi-Dirac integral and
$\tilde \mu_i$  for the effective chemical potential, 
$\mu_i=\tilde \mu_i + m_i c^2 +U_i$, with the mean-field potential
$U_i=\partial \epsilon_{g}/\partial \rho_{gi}$, and 
$\hbar^2/2m^*_i=\partial \epsilon_{g}/\partial \tau_{gi}$ 
gives the proton and neutron effective masses.
 
The phase diagram of homogeneous matter is known to present a complex 
phenomenology with temperature-dependent 1st and 2nd order phase transitions.
This means that for certain values of $(T,\mu_n,\mu_p)$ up to three solutions
exist and each can be in principle  put in equilibrium with the clusterized 
counterpart.
Among them, the equilibrium solution will be the one minimizing the 
thermodynamical potential.  

The clusterized component is regarded as a non-interacting gas of clusters 
of size $A_i>2$ and isospin $I_i=N_i-Z_i$ by
assuming that nuclear interactions are entirely exhausted by clusterization
or can be recasted in the form of in-medium modified cluster functionals.

The corresponding partition function writes
in the canonical ensemble,
\begin{equation}
{\cal Z}^{A>1}_{\beta,\mu_3}(A_0)=\sum_{\{n_A\}}\prod_{A>1}
{\frac{\omega_{A,\mu_3}^{n_A}}{n_A!}} 
\label{cano}
\end{equation}
To simplify the calculation of the partition sum, a saddle point 
approximation is made on the $I$ direction and only the most probable 
isotope $\bar{I}(A)$ for each size $A$ is retained.
The resulting expression for the partition sum of a cluster of size $A$ is:
\begin{eqnarray}
\omega_{A,\mu_3} =  \frac 12 \sqrt{2\pi\sigma^2_A}  V_F  
\left ( \frac{2\pi A m_0}{\beta h^2}\right )^{3/2}     
\nonumber \\
\exp-\left ( \beta  F^\beta_{A,\bar{I}} \right )
\exp (\beta\mu_3\bar{I}) \label{saddle}
\end{eqnarray}
where the most probable isotopic composition $\bar{I}$ of a cluster 
of size $A$ depends on the temperature according to,
\begin{equation}
\mu_3=\frac{\partial  F^\beta_{A,I} }{\partial I}|_{I=\bar{I}},
\end{equation}
and the associated isotopic dispersion is given by
\begin{equation}
\frac{1}{\sigma^2_A}=\beta \frac{\partial^2 F^\beta_{A,I}}{\partial I^2}|_{I=\bar{I}}.
\end{equation}

When expressing the partition functions we have introduced 
the isoscalar and isovector chemical potentials
$\mu=(\mu_n+\mu_p)/2$ and, respectively, $\mu_3=(\mu_n-\mu_p)/2$,
and we have neglected the difference between the proton and the neutron
bare mass, $m_n \approx m_p \approx m_0$.

$n_A=\sum_I n_{IA}$ is the occupation number of size $A$ where the sum 
is restricted to combinations   $\{n_A\}\equiv \{n_2,\dots,n_{A_0} \}$
satisfying the canonical constraint,
\begin{equation}
\sum_{A=2}^{A_0} A n_A=A_0. 
\label{canoconstr}
\end{equation}

$A_0$ is a big number corresponding typically to a few hundreds of 
Wigner-Seitz cells, and convergence is checked with respect to an 
increase of $A_0$ to insure that the 
thermodynamical limit is attained.
In eq.(\ref{saddle}) $V_F$ is the free volume associated 
to the cluster center of mass,
given by:
\begin{equation}
V_{F}(A)=\left[V - \frac{A_0-A}{\langle\rho_0\rangle_\beta}\right]
\left[ 1-\frac{\rho_g}{\rho_0^0}\right] ,
\end{equation}
where $<\rho_0>_\beta$ and $\rho_0^0$ stand for the average cluster density and
saturation density of symmetric matter. Different prescriptions are proposed
in Refs. \cite{hempel_npa2010,noi_2010} 
and the results are not very sensitive to the detailed approximation
employed.

One of the most important quantities is the cluster free energy 
$F^{\beta}_{A,I}$. 
In its most general case it writes \cite{ens-ineq}:
\begin{equation}
F^\beta_{A,I}=E_{A,I}+\langle E^*_{A,I} \rangle_\beta - TS^\beta_{A,I}, 
\end{equation}
where $E_{A,I}$ is the ground-state energy,
$\langle E^*_{A,I} \rangle_\beta$ is the average cluster excitation energy,
$S^\beta_{A,I}$ is the entropy.

In view of a consistent description of the nucleon and cluster gases, we
adopt for the cluster ground state energy functional the parameterization 
proposed by Danielewicz and Lee \cite{danielewicz_npa818} 
who provide for the parameters values fitted on Hartree-Fock calculations 
with a variety of Skyrme-interactions.
Additionally accounting for electron screening in the Wigner-Seitz 
approximation, the functional writes:
\begin{eqnarray}
E_{A,I}(\rho_e)=a_v A-a_s A^{2/3} -\frac{a_a(A)}{A} I^2
\nonumber \\
-a_c f_{WS}(A,I,\rho_e) \frac{\left (A-I\right )^2}{4A^{1/3}},
\label{eq:ldm_danielewicz}
\end{eqnarray}
 with
\begin{eqnarray}
a_a(A)=\frac{a_v^a}{1+a_v^a/(a_s^a A^{1/3})},
\end{eqnarray}
and
\begin{eqnarray}
f_{WS}(A,I,\rho_e)=1-\frac32 \left(  \frac{\rho_e}{\rho_{0p}(A,I)} \right)^{1/3}+
\frac12 \left(\frac{\rho_e}{\rho_{0p}(A,I)} \right), \nonumber \\
\end{eqnarray}
with $\rho_e$ and $\rho_{0p}(A,I)=(Y_p)_{cl}\rho_0(A,I)$ standing for the 
electron density and, respectively, proton density inside the cluster. 
Here, $\rho_0(A,I)$ is the saturation density corresponding to 
the isospin asymmetry in the cluster bulk,
\begin{equation}
\rho_0(A,I)=\rho_0^0 \left( 1-\frac{3 L \delta_{cl}^2}{K+K_{sym} \delta_{cl}^2}
\right),
\end{equation}
meaning that we account for fragment compressibility in agreement with 
microscopic findings \cite{panagiota}.
In principle in this expression $\delta_{cl}=1-2(Y_p)_{cl}$ should represent the 
isospin asymmetry in the cluster bulk, which is here approximated for 
simplicity to the average cluster asymmetry $\delta_{cl}=I/A$. 

A consistent calculation of the cluster excitation energy 
$\langle E^*_{A,I} \rangle_\beta$ and level density $\exp( S^\beta_{A,I})$
with the same Skyrme functional used for the energy is beyond the scope 
of this work.
We will consider a simple Fermi-gas expression as often employed in the 
literature, but limit the study to temperatures low enough that the 
ambiguity associated to this inconsistency plays a negligible role.

The different physical quantities are calculated as a weighted average of 
the cluster $x_{cl}=A_0/A_{tot}$ and unbound $(1-x_{cl})=A_{free}/A_{tot}$ 
nucleons component.
For instance, accounting for the excluded volume, the total baryonic density 
reads:
\begin{equation}
\rho=\frac{A_0}{V}+\rho_{g}
\left ( 1-\frac{A_{0}}{V<\rho_0>_\beta}\right ). 
\label{excl}
\end{equation}

\subsection{The different effective interactions}

\begin{table*}
\caption{Bulk nuclear properties for different Skyrme interactions as
given in Ref. \cite{danielewicz_npa818}}
\begin{tabular}{llllllllll}
\hline
NN-potential & $\rho_0^0$   & $K$  & $L$   & $K_{sym}$ & 
$a_v$  & $a_s$ & $a_v^a$ & $a_s^a$ & $a_c$\\
$~$          & (fm$^{-3}$)& (MeV) & (MeV) & (MeV) & (MeV)   & (MeV)  & 
(MeV) & (MeV) & (MeV)   \\
\hline\noalign{\smallskip}
SLY4         & 0.1595    & 230.0   & 46.0  & -119.8 & 
15.97 & 18.24 & 32.00   & 16.60 & 0.69 \\
SGI          & 0.1544     & 261.8   & 63.9  & -52.0 & 
15.89 & 17.48 & 28.33   & 12.76 & 0.69 \\
SkI3         & 0.1577    & 258.2  & 100.5 & 73.0  & 
15.98 & 17.77  & 34.83   & 12.77 & 0.69 \\
LNS          & 0.1746    & 210.8 & 61.5  & -127.4 & 
15.31 & 15.77  & 33.43   & 14.10  & 0.69 \\  
\noalign{\smallskip}\hline
\end{tabular}
\label{table:NNparam}
\end{table*}

To study the sensitivity to the symmetry energy, four effective interactions 
have been considered: SLY4 \cite{sly4}, 
SGI \cite{sgi}, SkI3 \cite{ski3} and LNS \cite{lns}.
Their properties in terms of saturation density for symmetric matter $\rho_0^0$,
compression modulus $K$, 
slope of the symmetry energy  $L$ 
and curvature of the symmetry energy $K_{sym}$ 
around $\rho_0^0$ are listed in Table \ref{table:NNparam}.
They have been chosen such as to behave similarly in the
isoscalar direction and different in the isovector direction. In this way, 
the observed differences in the predictions will be straightforwardly 
associated to the differences in the isovector or symmetry behavior.

The similarity of the isoscalar features is indicated by the relatively
narrow range the values of $K$ spanned, in agreement with present 
constraints from collective modes and heavy ion collisions.
The different isovector features are illustrated by the different values of
the slope $L$ and the curvature $K_{sym}$ of the symmetry energy around
$\rho_0^0$. The broad ranges, $46 \leq L \leq 100.5$ MeV and
$-127.4 \leq K_{sym} \leq 73$ MeV, are suggestive of how little the present
available experimental data still constrain the EOS.

\begin{figure*}
\resizebox{0.99\textwidth}{!}{
\includegraphics{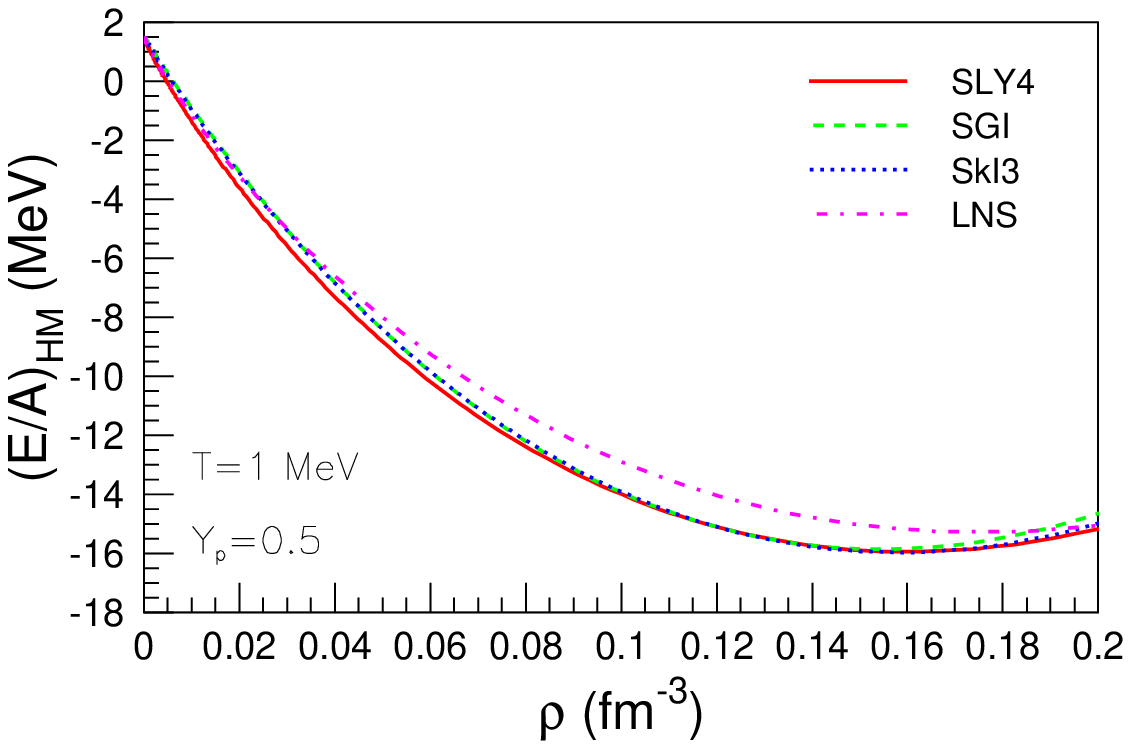}
\includegraphics{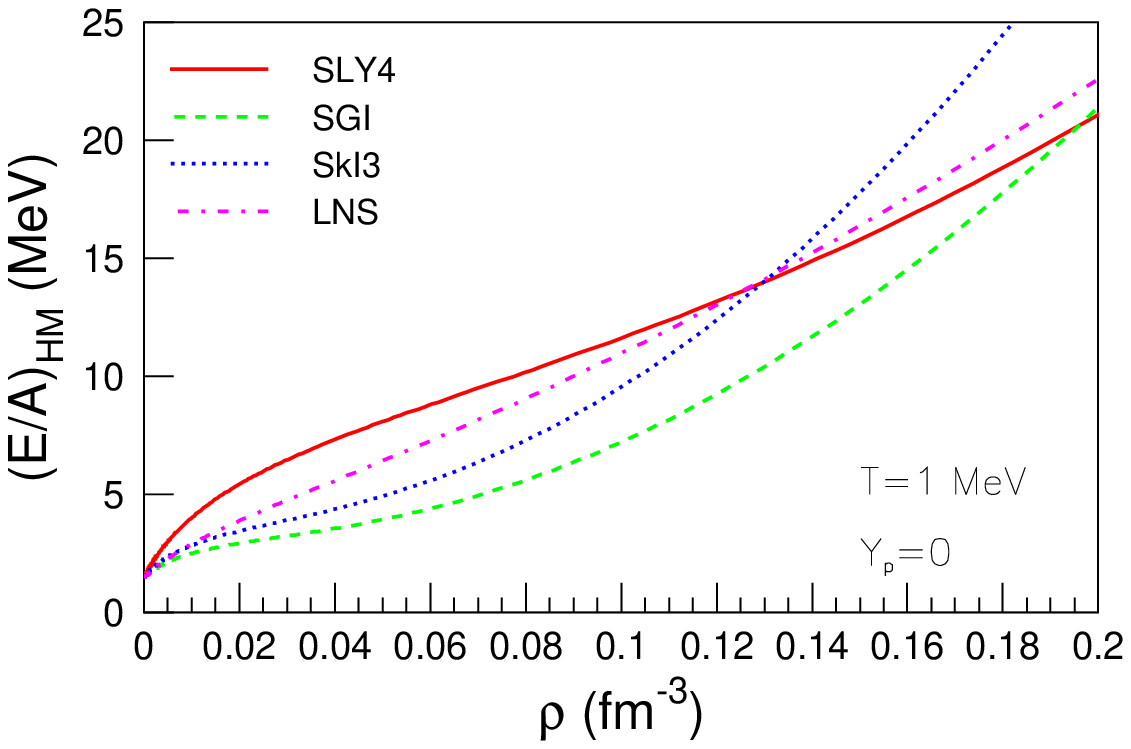}
}
\resizebox{0.99\textwidth}{!}{
\includegraphics{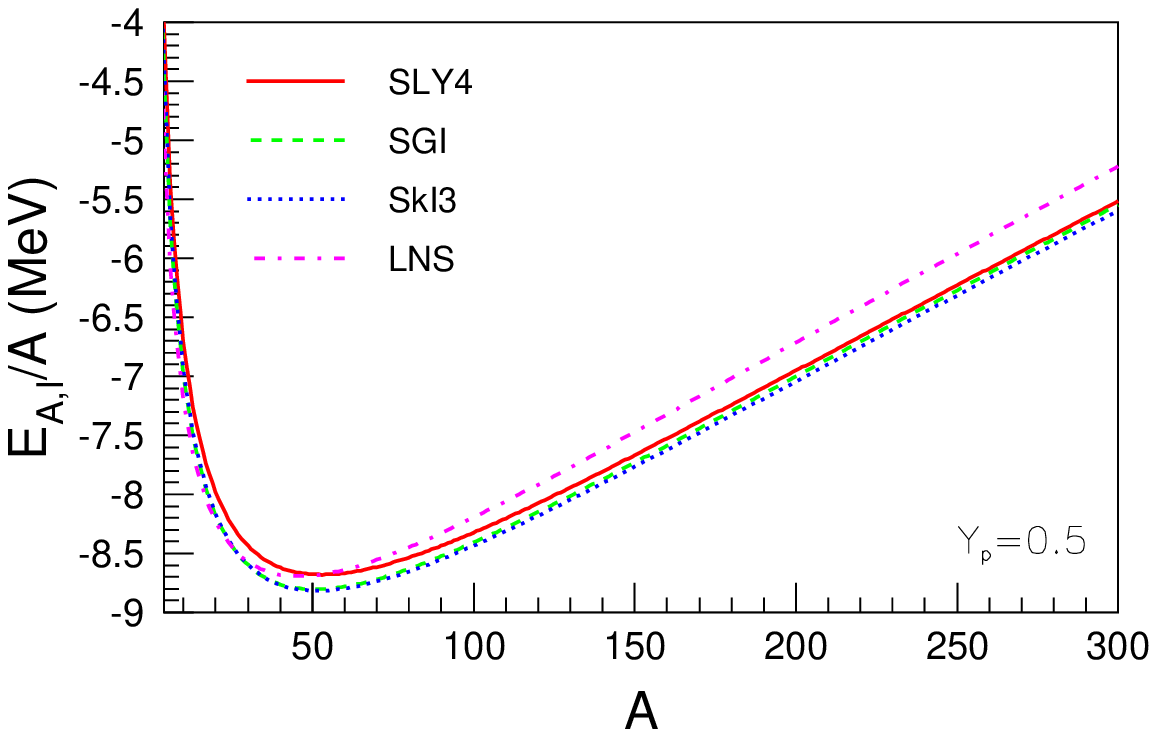}
\includegraphics{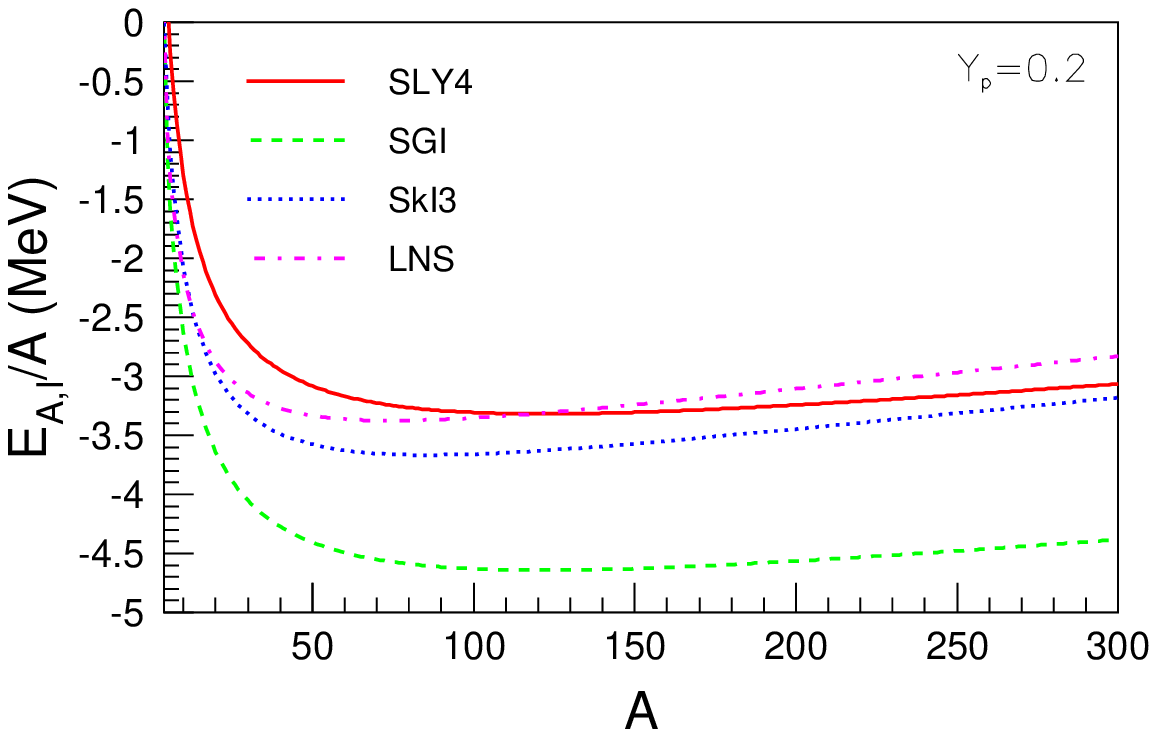}
}
\caption{
Energy per baryon versus $\rho$ (top) and versus $A$ (bottom) 
for different NN-interaction parameterisations
in the case of homogeneous matter (at $T$=1 MeV) and, respectively, nuclei
as predicted by the self-consistent mean-field theory, corresponding to 
different proton fractions as mentioned on each panel. 
}
\label{fig:ingredients}
\end{figure*}

The evolution with density of the energy per baryon provided by
the different effective interactions is plotted in the upper panels 
of Fig. \ref{fig:ingredients} for
symmetric matter ($Y_p=0.5$) (left) and 
neutron matter ($Y_p \equiv\rho_p/\rho$=0) (right)
at the arbitrary value of 1 MeV temperature for which most of the NSE 
calculations shown in this paper are performed (see below).
For symmetric matter, the three EOS characterized by the closest values of $K$,
that is SLY4, SGI and SkI3, lead
to $E/A(\rho)$-curves that sit one on the top of the other while a small 
shift is obtained for LNS, due to the slightly too high saturation density 
presented by this parametrization.
On the contrary, the energetics of the neutron-pure matter shows over the whole
density range a significant sensitivity to the effective interaction.

The lower panels in Fig. \ref{fig:ingredients} depict the evolution with the
cluster mass number of the binding energy per nucleon as predicted by 
Eq. (\ref{eq:ldm_danielewicz}) for the four considered effective interactions
and two arbitrary values of the proton fraction: 0.5 (left) and 0.2 (right).
The values of the LDM parameters are taken from Ref.
\cite{danielewicz_npa818} and are listed in Table \ref{table:NNparam}.
Little sensitivity to the effective interaction is shown by isospin-symmetric
clusters, reflecting the good constraint on the isoscalar equation of state,  
while the opposite holds for the neutron-rich ones.
Specifically one can see that the cluster energetics does not reflect the 
behavior of the EOS at $\rho_0^0$ but rather at a lower density, 
where the difference between the different interactions is important. 
This is essentially due to the surface term, and shows the importance of 
employing parametrizations for the energetics of the clusters, 
which are consistent with the modelization of the uniform matter in the core. 

As we will show, $\beta$-equilibrium matter at low temperature is 
extremely neutron rich. For this reason we expect that an important 
sensitivity on the effective interaction, 
and the underlying symmetry energy, will persist even in clusterized matter. 

\subsection{Bulk in-medium effects}
\label{deltaEbulk}

\begin{figure*}
\resizebox{0.99\textwidth}{!}{
\includegraphics{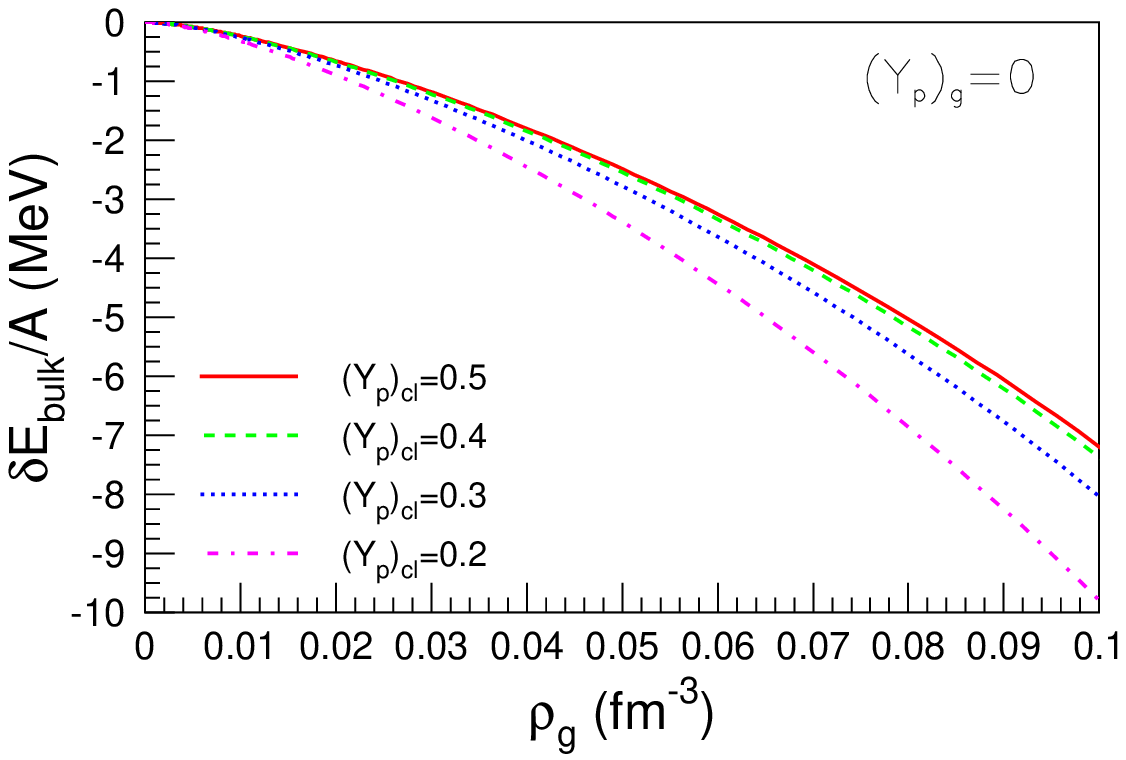}
\includegraphics{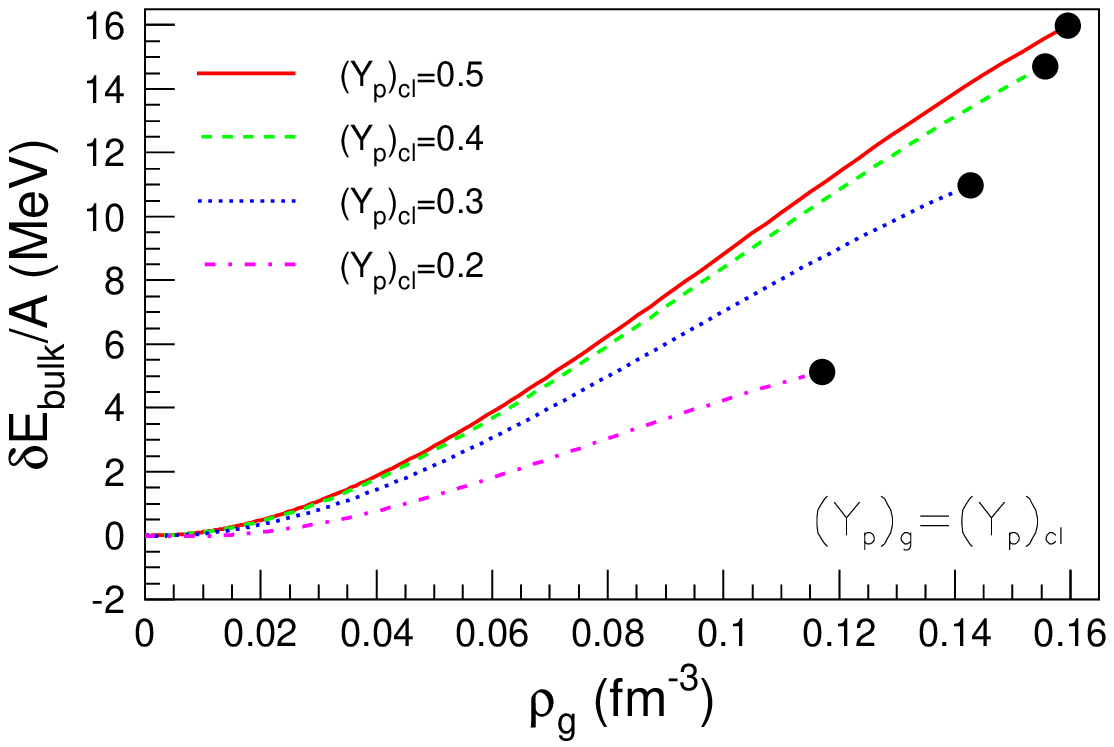}
}
\caption{$\delta E_{bulk}/A = -\epsilon(\rho_{gn},\rho_{gp})/\rho_0(A,I)$ 
versus $\rho_g$  
for two different gas compositions: $(Y_{p})_g=0$ (pure neutron gas) 
and $(Y_{p})_g=(Y_p)_{cl}$ (same asymmetry) for different values of
$(Y_p)_{cl}$ as mentioned on each panel. 
The effective interaction is SLY4. 
The ending points of $\delta E_{bulk}/A$  
mark the densities corresponding to cluster dissolution in the dense medium.
}
\label{fig:deltaEbulk}       
\end{figure*}

At finite temperature the system is not periodic, and the concept of 
Wigner-Seitz cell is not fully meaningful. However a cluster-dependent 
Wigner-Seitz volume can still be defined as the volume neutralizing 
the cluster charge as:
\begin{equation}
V_{WS}(A)=\frac {V}{A_0} A,
\end{equation}
while the average Wigner-Seitz volume is given as a function of the average 
cluster mass in the cell, 
$\langle V_{WS}\rangle_\beta=V\langle A_{cl}\rangle_\beta/A_0$.
This definition converges to the standard definition at $T=0$.
Using Eqs. (\ref{eq:egas}), (\ref{eq:ldm_danielewicz}), (\ref{excl}), 
the total energy inside a single Wigner-Seitz cell containing a cluster 
$(A,I)$ and unbound neutrons and protons at 
a density $\rho_g=\rho_{gn}+\rho_{gp}$ and an asymmetry
$\delta_g=(\rho_{gn}-\rho_{gp})/\rho_g$, reads:
\begin{equation}
E_{WS}=E_{A,I}(\rho_e)+\epsilon(\rho_{g},\delta_{g})\left (V_{WS}-
\frac{A}{\rho_0(A,I)}\right )
\end{equation}
where the energy functional of the unbound particles corresponds to 
the energy density of an infinite homogeneous system at a density 
$\rho_{gn},\rho_{gp}$ calculated in the mean-field approximation. 

We have derived this expression making use of the classical excluded volume 
concept. We can alternatively write the Wigner-Seitz energy in terms of 
the complete Skyrme energy density functional
$\epsilon\left[\{\rho_i,\tau_i\}\right]$,  $ i=n,p $ 
including gradient and non-local terms as
\begin{eqnarray}
E_{WS}&=& \int_{V_{WS}}  \epsilon \left[ \{\rho_i(r),\tau_i(r)\} \right]d^3r 
\nonumber \\
&=& \int_{A/\rho_0}  \epsilon \left[ \{\rho_i(r),\tau_i(r)\} \right]d^3r
\nonumber \\
&+&\int_{V_{WS}-A/\rho_0}  \epsilon \left[ \{\rho_i(r),\tau_i(r)\} \right]d^3r 
\nonumber \\
&=& E_{A,I}(\rho_e)+\epsilon(\rho_{gn},\rho_{gp})
\left (V_{WS}-\frac{A}{\rho_0(A,I)}\right ) \nonumber \\
&+& \delta E_{surf}
\label{medium}
\end{eqnarray}
where $\delta E_{surf}$ represents the interface energy between the cluster 
and the gas, and is expected to scale as the cluster surface 
$\propto R^2\propto \rho_0(A,I)^{-2/3}A^{2/3}$. 
We can consider this interface energy as an in-medium modification of the 
cluster energy functional and write
\begin{equation}
E_{WS}=  E_{A,I}^m(\rho_e,\rho_{gn},\rho_{gp})+\epsilon(\rho_{gn},\rho_{gp})V_{WS} 
\end{equation}
  with
\begin{equation}
 E_{A,I}^m(\rho_e,\rho_n,\rho_p)= E_{A,I}(\rho_e) +\delta E_{bulk} +\delta E_{surf}
\end{equation}
where the bulk binding energy shift is given by
\begin{equation}
\delta E_{bulk}=-\frac{\epsilon(\rho_{g},\delta_{g})}{\rho_0(A,I)}A
\end{equation}
The physical origin of this in-medium modification is easy to understand.
In the mean-field approximation, the single particle states $|i>$ 
composing the nuclear gas are plane waves delocalized over the whole volume, 
and thus partially contributing to the local density 
 of the cluster, $\rho(r)=\sum_{i=1}^{A+A_{free}} <r|i>^2$.
Since the energy minimization in the Wigner-Seitz cell leads to a 
bulk cluster density  equal to the saturation density at the corresponding 
isospin asymmetry \cite{panagiota}, 
the bulk cluster energy is reduced 
by the contribution of the continuum states. 
This argument shows that the excluded volume mechanism correctly accounts 
for the bulk part of the in-medium binding energy correction, at least 
in the local density approximation.

To quantitatively understand how important the correction 
$\epsilon/\rho_0$ is, we plot in 
Fig. \ref{fig:deltaEbulk} the evolution of this quantity as a function of
gas density for two representative cases of a
pure neutron gas (left) and a gas asymmetry equal to the cluster one (right).
The effective interaction considered here is SLY4.

Imposing the gas asymmetry to be equal to the cluster asymmetry, 
amounts to disregard isospin effects (isospin fractionation) in the 
thermodynamical equilibrium.
In this case we recover the well known result that the cluster energy is 
reduced by the presence of the surrounding medium, leading to the
dissolution of clusters at the critical Mott density \cite{roepke}. 
This density can be defined as the density corresponding to vanishing 
bulk binding, and is given by the ending point of each curve in Fig.  
\ref{fig:deltaEbulk}. We can see that this critical density 
monotonically decreases with increasing cluster asymmetry.

In the case of stellar matter at $\beta$-equilibrium the fractionation effect 
cannot be neglected, and the gas is systematically more neutron-rich than 
the clusters. In the simplified $T=0$ case of the neutron star crust,
unbound particles are uniquely constituted of neutrons 
\cite{negele_vautherin}.
The limiting case $(Y_p)_g=(\delta_g-1)/2=0$ is thus closer to the physical 
condition of the stellar environment. In this case the trend is reversed. 
The unbound component being strongly asymmetric, 
the bound part of matter associated to the cluster is more symmetric, 
as a part of the neutron single-particle states are continuum states 
belonging to the gas.
Being effectively more symmetric than if the gas was not there, 
the cluster is more bound.
This simple mechanism explains why clusters can survive in environment 
extremely neutron rich as neutron star crusts.  

As a first approximation, we can consider that the binding energy shift 
implied by the excluded volume mechanism is the dominant in-medium effect. 
For this reason, in the next section we present results of the extended NSE 
model neglecting the in-medium surface correction $\delta E_{surf}$.

\section{Results at $\beta-$equilibrium at finite temperature}
\label{section:NSE-betaeq}

\begin{figure}
\resizebox{0.99\columnwidth}{!}{
\includegraphics{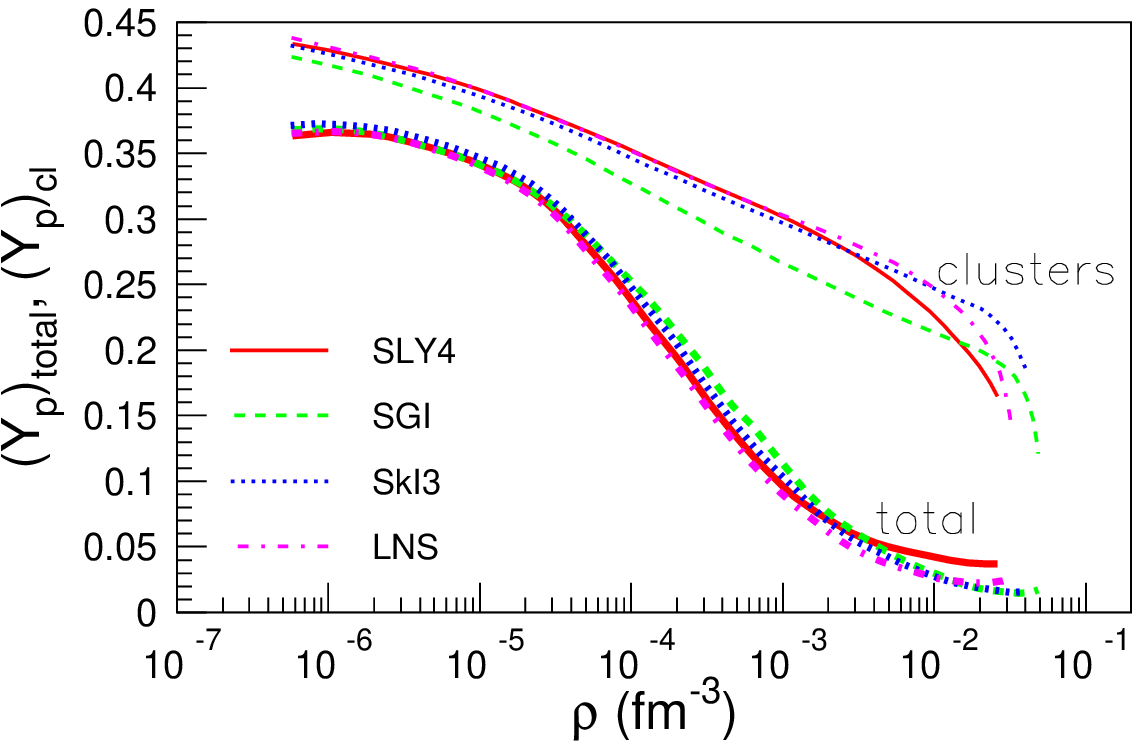}
}
\resizebox{0.99\columnwidth}{!}{
\includegraphics{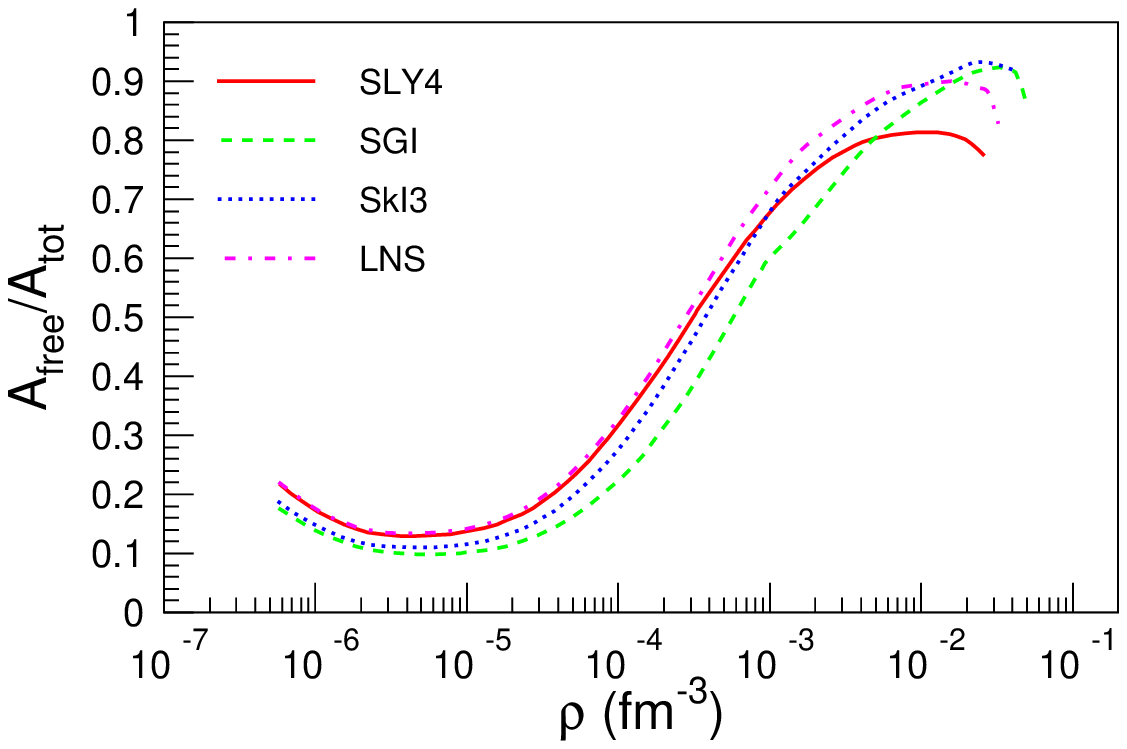}
}
\resizebox{0.99\columnwidth}{!}{
\includegraphics{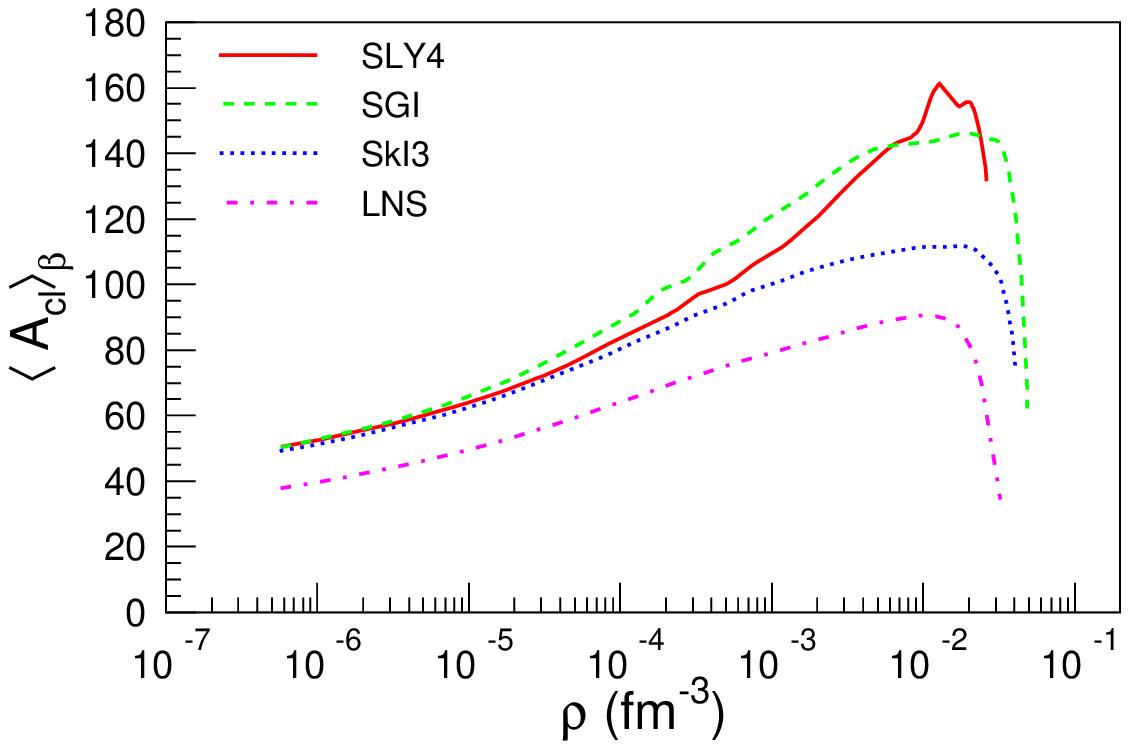}
}
\caption{$T$=1 MeV; Average cluster properties (size and proton fraction), 
total $Y_p$ and percentage of unbound nucleons as a function of 
total baryonic density along the $\beta-$equilibrium path. 
}
\label{fig:cluster_prop_T=1}
\end{figure}
 
Properties of dilute clusterized baryonic matter 
relevant for CCSN and PNS
as total baryonic energy and entropy per baryon, pressure,
relative amounts of bound and unbound matter, size of the most probable
cluster, etc.
are often plotted in NSE as a function of total baryonic density along
constant proton-fraction paths. This choice is mostly motivated
by the fact that in the core collapse evolution before and after bounce a very 
wide interval of temperatures, densities and proton fraction is explored, 
and the equation of state is needed in this three-dimensional space.

If however we limit ourselves to the low temperature case relevant for (proto)
neutron star crusts and the first steps of the CCSN dynamics, 
the proton fraction at each baryonic density is determined by the neutrinoless 
$\beta$ - equilibrium condition, and the problem is reduced to a one-dimensional
space. 

The upper panel of Fig. \ref{fig:cluster_prop_T=1} illustrates
the overall proton fraction, as well as the average proton fraction 
inside the clusters, as a function of density for 
$T$=1 MeV at $\beta$-equilibrium.  
The four effective interactions discussed in the previous 
section have been considered.
$Y_p(\rho)$ shows a monotonic decrease from a value slightly below 0.4 at
$\rho \approx 10^{-7}$ fm$^{-3}$ to almost zero at 
$\rho>5 \cdot 10^{-2}$ fm$^{-3}$, in agreement with the pioneering work of 
Negele and Vautherin \cite{negele_vautherin}.
The $Y_p(\rho)$-curve seems to be largely independent of the 
effective interaction.  
This can be understood taking into account that $Y_p=Y_e$
is determined, for the neutrino free steaming regime here assumed, 
by the $\beta$-equilibrium relation 
where the ideal character of the electron gas dominates over
the details of the nuclear interactions.

In contrast with this, the relative number of unbound nucleons 
$A_{free}/A_{tot}=(1-x_{cl})$ presented in the medium panel
indicates a possible correlation between
$L$ and the crust-core transition width.

Roughly speaking, over the considered density range $A_{free}/A_{tot} (\rho)$
increases from 0 to 1. This confirms that at low densities matter is chiefly
made out of clusters while at high densities it rather consists of uniform
matter in strong interaction. 
A closer look reveals however a non-monotonic behavior at the extreme densities.
The initial decrease of the gas component is a finite temperature effect:
at low density matter is almost symmetric and clusters
are below the neutron-drip line. 
The population of continuum states is thus solely due 
to the occupation of single-particle states above the Fermi energy 
due to the finite temperature. The total number of nucleons in the 
continuum is proportional to the available volume, and thus decreases 
with increasing density. 
As density increases, the system becomes globally more and more neutron-rich.
When the neutron drip-line is overcome, an extra contribution of 
unbound neutrons in the continuum states appears, 
which monotonically increases with increasing density
independent of the temperature. 

$\langle (Y_{p})_{cl}\rangle_\beta (\rho)$ decreases monotonically as 
a consequence of the monotonic decrease of $Y_p(\rho)$ and, irrespective 
the density, $\langle (Y_{p})_{cl}\rangle_\beta >Y_p $. 
This means, as expected, that the 
dense matter is always more symmetric than neutron matter, 
consistent with the zero temperature limit
of exclusive neutron drip. 
The lower panel of Fig. \ref{fig:cluster_prop_T=1} gives the 
additional information of the average cluster mass.  
The generic shape of $\langle A_{cl} \rangle_\beta (\rho)$
shows a gentle increase over several orders of magnitude in $\rho$ and, for
$\rho> 10^{-2}$ fm$^{-3}$, a sudden fall. 
While the average cluster increase is the NSE replica of pastas where the dense
phase expands with increasing the density, the fall is the consequence of the
very small number of protons available at high densities and which 
are essential for cluster formation. 

The decrease of cluster size due to decreasing proton fraction explains 
the high density decrease of  $A_{free}/A_{tot} (\rho)$ observed above.
Indeed the reduced cluster charge implies a reduced Wigner-Seitz volume 
and an increased cluster volume, which tend to reduce the percentage of 
unbound nucleons.

The most important sensitivity to the EOS concerns the average cluster size.
This can be understood from the fact that at low proton fractions the 
size of the most stable cluster is strongly connected to the isovector 
properties of the effective interaction, 
as we have observed commenting Fig.\ref{fig:ingredients}.
 
\begin{figure*}
\resizebox{0.99\textwidth}{!}{
\includegraphics{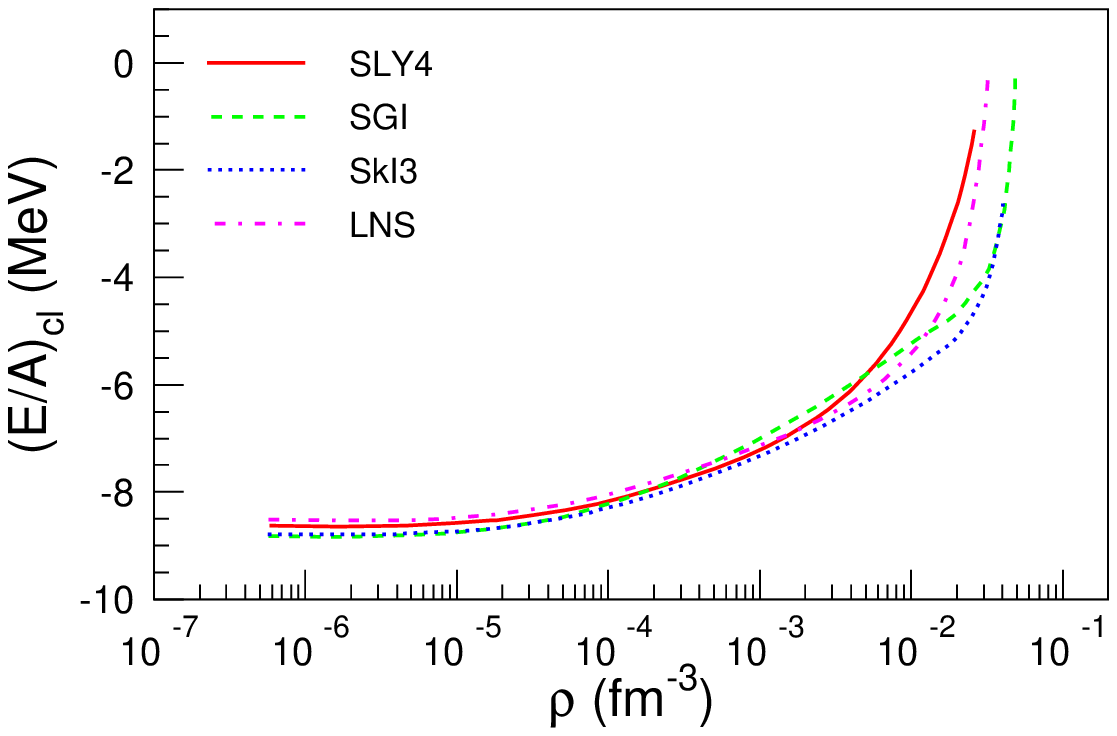}
\includegraphics{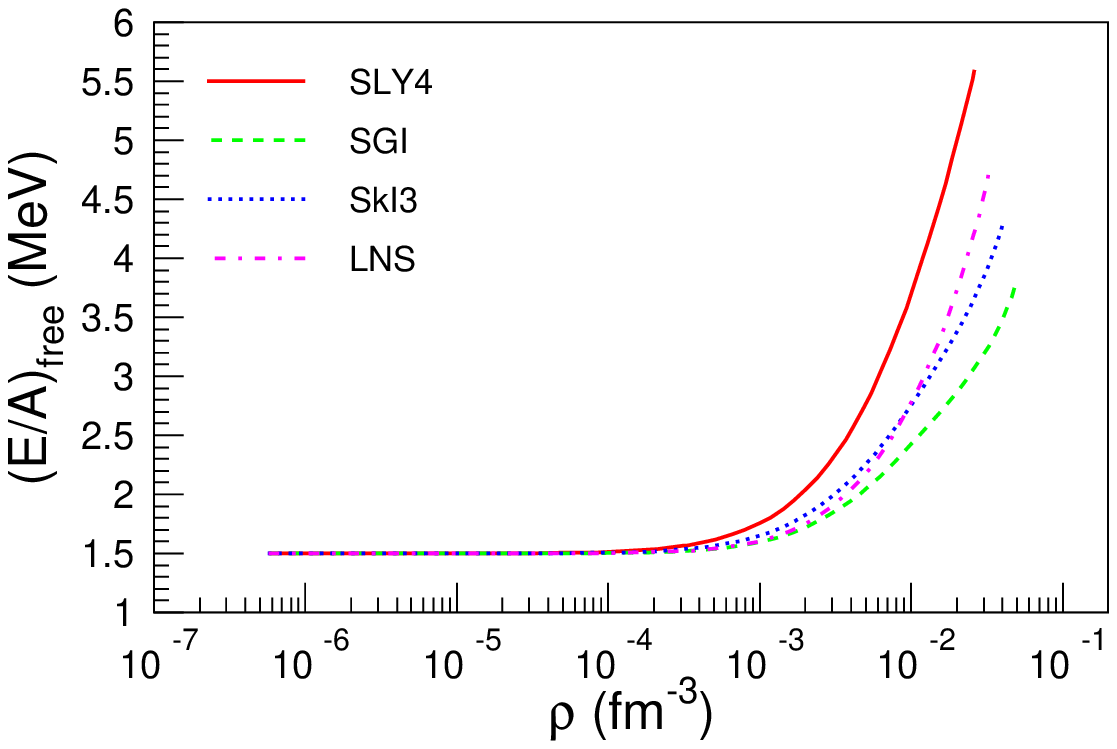}
}
\resizebox{0.99\textwidth}{!}{
\includegraphics{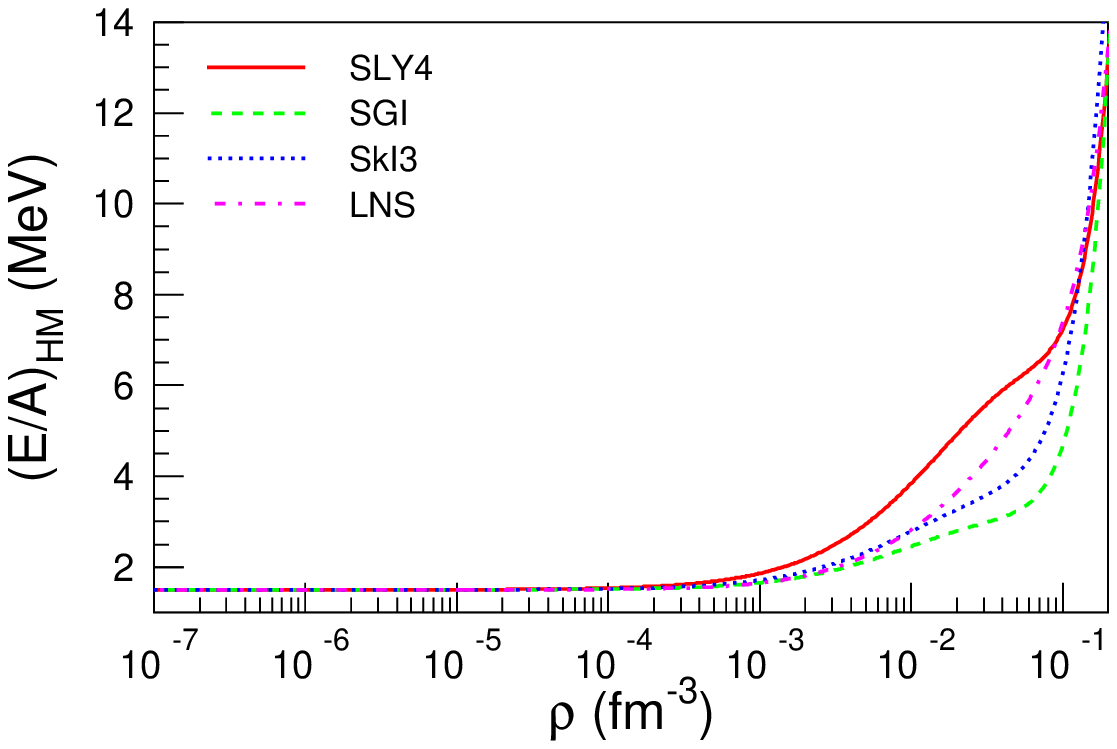}
\includegraphics{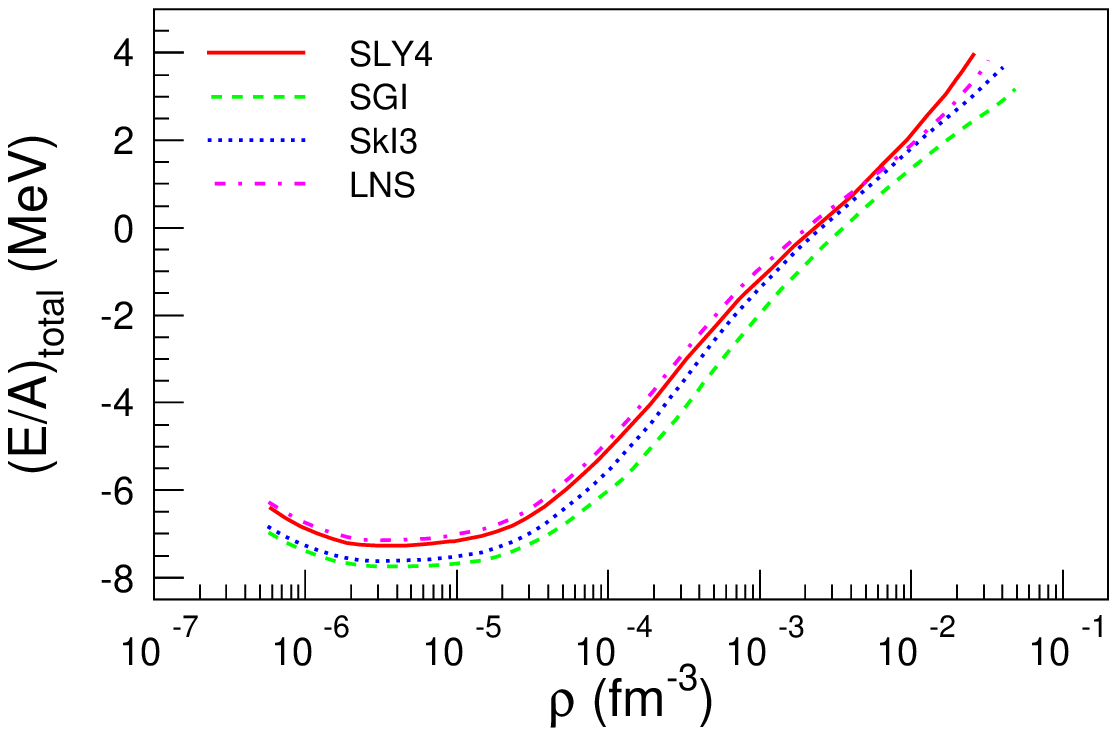}
}
\caption{$T$=1 MeV; Energy per baryon as a function of total baryonic 
density along the $\beta-$equilibrium path.
Upper panels: Clusterized (left) and unbound (right) sub-systems; 
Lower part: Total system (right) and pure homogeneous matter (left).
}
\label{fig:part_energies_T=1}
\end{figure*}

Fig. \ref{fig:part_energies_T=1} depicts the evolution with density
of the total baryonic energy per baryon corresponding to the 
total system and, respectively, the free and bound nucleons component 
separately.
For the sake of completeness also the energetics of pure homogeneous matter
at $T$=1 MeV and $\beta$-equilibrium is considered.
Comparing the behavior of uniform matter with the behavior of the 
total inhomogeneous system we can see that correctly accounting for 
the clusters is absolutely essential to understand the energetics of 
stellar matter, and no conclusion on the compact stars physics 
can be drawn based on the mean-field behavior of nuclear matter.
 
In particular, at very low densities ($\rho <10^{-5}$ fm$^{-3}$), 
where the unbound component is very dilute and
represents a small fraction of matter, the energetics of the system is
determined to a large extent by the energetics of clusters. 
In this regime, being not far from symmetry, the clusters are similar to 
terrestrial nuclei, are characterized by an energy per nucleon of the order 
of -8.5 MeV and all the different predictions coincide.
However we have already seen that, even at these low densities,
the matter composition as measured by  
$A_{free}/A_{tot}$  shows a dependence on the interaction.
This explains why, in contrast to the non-sensitivity of $(E/A)_{free}$,
a certain spread of the order of 1 MeV per nucleon is observed in the 
predictions for $(E/A)_{tot}$ even in this region of moderate isospin 
asymmetries where the effective interaction is well constrained.

In the density domain where the unbound component dominates 
($\rho> 10^{-3}$ fm$^{-3}$) the mixture is unbound, too.
The dispersion of $(E/A)_{tot}$ here is not larger than the one at low densities,
contrary to what one would have expected considering the behavior 
of $(E/A)_{free}$. This can be understood from the fact that the different 
energetics and different compositions are not correlated.

Globally speaking, we can say that the presence of clusters in 
dilute matter reduces the uncertainties due to our incomplete knowledge 
of the isovector equation of state.
Still a dispersion in the predictions is seen, and it is clear that it is 
essential to determine the isovector behavior at low density better than 
the present constraints in order to have a reliable model for the equation 
of state of stellar matter.

\section{In-medium surface effects}
\label{section:deltaEsurf}

Microscopic calculations \cite{douchin} indicate that surface properties 
of clusters are modified by the presence of an external medium. 
This means that, even if the dominant in-medium effect 
is accounted by the excluded volume mechanism, we should expect that the 
correcting term $\delta E_{surf}$ cannot be negliged in general.

From equation (\ref{medium}) we can deduce the expression of this in-medium 
binding energy shift 
in the framework of the density functional theory:
\begin{eqnarray}
\delta E_{surf}(A,I,\rho_{ng},\rho_{pg})=\int_{V_{WS}}  \epsilon 
\left[ \{\rho_i(r),\tau_i(r)\} \right]d^3r 
\nonumber \\
- E_{A,I}(\rho_e)-\epsilon(\rho_{gn},\rho_{gp})
\left (V_{WS}-\frac{A}{\rho_0(A,I)}\right ) 
\label{surf1}
\end{eqnarray}
In the local density approximation 
$ \epsilon \left[ \{\rho_i(r),\tau_i(r)\} \right]$ $\approx$ 
\newline
 $\epsilon ( \rho_n(r),\rho_p(r) )$, 
eq.(\ref{surf1}) can be easily solved if the density profiles
$\rho_i(r)$ are known. It is well known \cite{brack,treiner} 
that a variational estimation of the density profile leads to a 
good estimation of Hartree-Fock energies only if the fourth order
correction in the $\hbar$ expansion for the kinetic energy densities is 
included in the extended Thomas Fermi approach, or alternatively if 
adjustable parameters are fine-tuned. 
For this reason it was recently proposed in ref. \cite{panagiota} 
an alternative modelization where an analytical ansatz for the density 
profile is checked against Hartree-Fock calculations in the Wigner-Seitz 
cell, and the energy is calculated using the simpler local density 
approximation corresponding to the lowest 
$\hbar$ Thomas-Fermi order, 
that is neglecting the higher order gradient terms
in the kinetic energy density and effective mass. 

The density profile is given by the convolution between a flat 
$\rho_g$ and a rapidly falling distribution of 
Woods-Saxon type associated respectively to the gas and cluster density
\cite{panagiota}:
\begin{equation}
\rho_{i}(r)=\frac{\rho_{0i}-\rho_{gi}}{1+\exp\left( 
\frac{r-R_{i}}{a_i} \right)}+\rho_{gi}; ~~ i=n,p, \label{density}
\end{equation}
where the radius parameter is given in terms of the equivalent homogeneous 
sphere radius $R^{HS}$ as 
$R_i = R_i^{HS} ( 1 - \pi^2/3 ({a_i}/{R_i^{HS}})^2 )$ , 
$\rho_{0i} = ({A\pm I })\rho_0(A,I)/(2A)$ are the partial saturation 
densities calculated for the cluster asymmetry, 
and the diffuseness parameters have a quadratic dependence on the 
bulk asymmetry, $a_i=\alpha_i +\beta_i (1-2(Y_p)_{cl})^2$. 
For details, see Ref. \cite{panagiota}. 
 
Even if the density profile (\ref{density}) gives an excellent reproduction 
of the Hartree-Fock calculation, the associated energy calculated 
in the local density approximation deviates from the microscopic result 
\cite{panagiota}.
This is due to the absence of spin-orbit and non-local terms in the 
kinetic energy density in the LDA. These energy terms are not affected by 
the external medium, and it was shown \cite{panagiota} that the deviation 
of the LDA with respect to HF is constant with the gas density.
This means that we can calculate the in-medium surface correction from the LDA 
approximation, provided the vacuum energy is consistently derived 
from the same approximation:
\begin{eqnarray}
\delta E_{surf}(A,I,\rho_{ng},\rho_{pg})=\int_{V_{WS}}  \epsilon
\left(\rho_n(r),\rho_p(r)\right)d^3r
\nonumber \\
- \int_{V_{WS}}  \epsilon_{\rho_g=0} \left(\rho_n(r),\rho_p(r)\right)d^3r \nonumber \\
- \epsilon(\rho_{gn},\rho_{gp})\left (V_{WS}-\frac{A}{\rho_0(A,I)}\right ) 
\label{surf2}
\end{eqnarray}
where $ \epsilon_{\rho_g=0}$ is obtained putting $\rho_{gn}=\rho_{gp}=0$ in 
eq. (\ref{density}).

\subsection{Study of $\delta E_{surf}$}

\begin{figure*}
\resizebox{0.99\textwidth}{!}{
\includegraphics{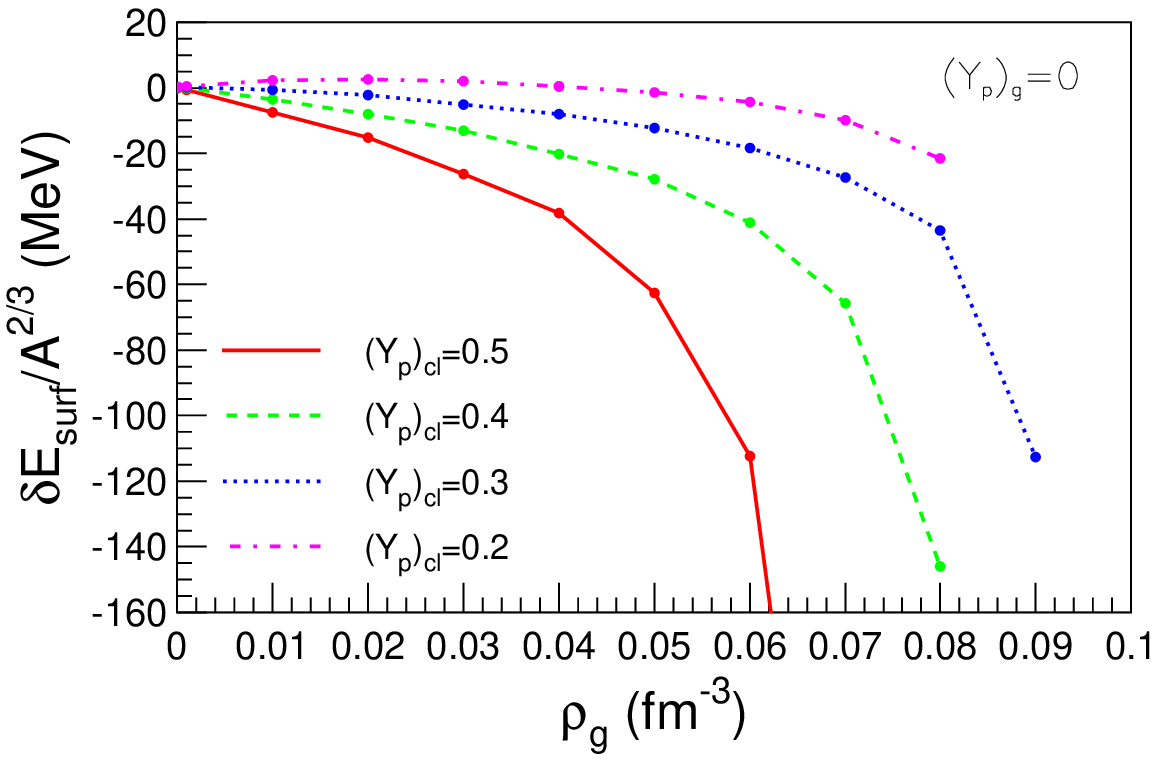}
\includegraphics{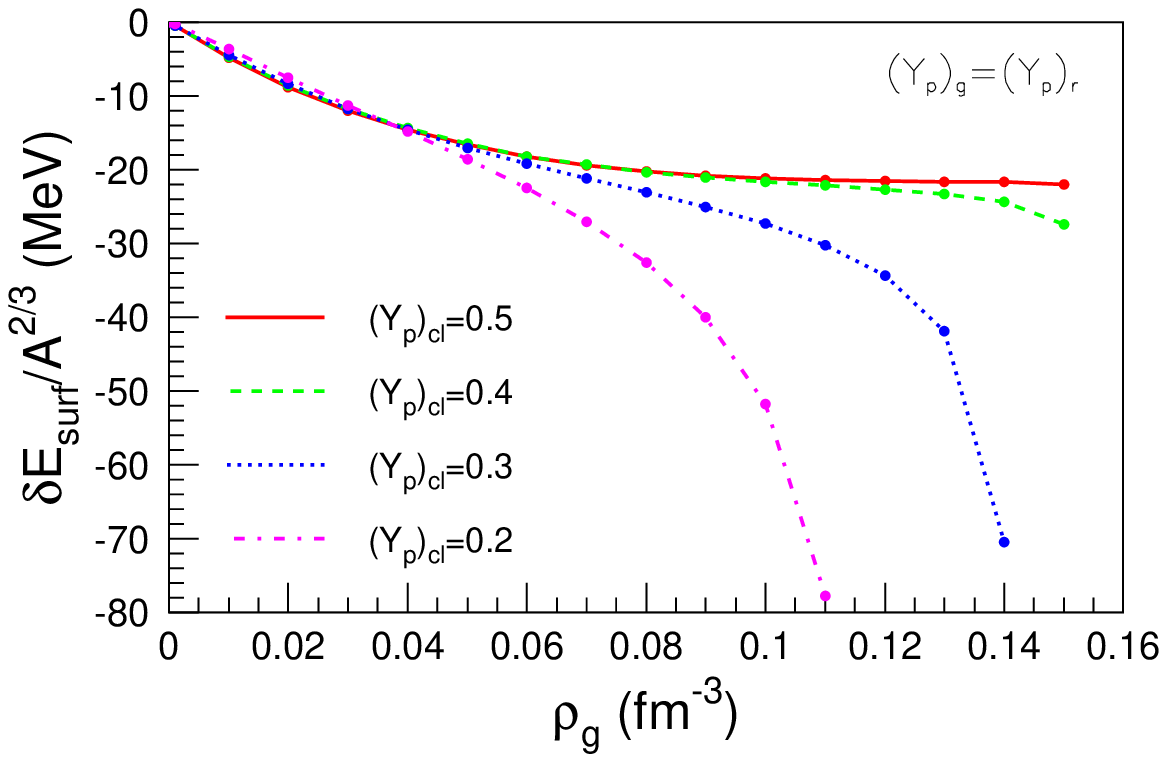}
}
\caption{$\delta E_{surf}/A^{2/3}$ versus $\rho_g$ 
for two different gas compositions: $(Y_{p})_g=0$ (pure neutron gas) 
and $(Y_{p})_g=(Y_p)_{cl}$ (same asymmetry)  
for different values of $(Y_p)_{cl}$  as mentioned on each panel.
The effective interaction is SLY4.
}
\label{fig:deltaEsurf}
\end{figure*}

Fig. \ref{fig:deltaEsurf} illustrates the surface tension, 
defined as the scaled in-medium modification of surface
energy  $\delta E_{surf}/A^{2/3}$ as a function of gas density $\rho_g$  
for two particular cases:  $(Y_{p})_g=0$   (pure neutron gas) 
and  $(Y_{p})_g=(Y_p)_{cl}$ (gas asymmetry is identical to the cluster one).
The calculation was done varying the cluster size 
and isospin over a very large domain of $N$ and $Z$ covering the 
whole periodic table well beyond the neutron dripline. 
The perfect scaling with $A^{2/3}$ observed shows 
that indeed the residual in-medium binding energy shift is a surface effect.

In symmetric matter (right side, curve labeled $(Y_p)_{cl}=0.5$) the surface 
energy is reduced in the medium and vanishes at normal density, 
being already negligible around $\rho_g\approx \rho_0^0/2$.
This is due to the compensation of the surface energy due to the finite size, 
by the attractive interaction with the surrounding gas, which becomes 
indistinguishable from the cluster bulk at $\rho_g = \rho_0^0$.
Increasing the isospin asymmetry and ignoring isospin fractionation effects 
(right side), the bulk density reduces and the density of the neutron gas 
can overcome the neutron density inside the cluster. This bubble-like effect 
leads to a negative surface energy, and is at the origin of the appearance 
of pasta phases in dense matter.
 
In the pure neutron gas case (left part), the behavior is opposite, 
similar to what we have observed for the bulk energy shift.
Indeed, the interface interaction with the surrounding gas becomes less 
attractive if the cluster  is more  asymmetric, leading to a decrease of 
the binding and therefore an increase of the surface  energy
with increasing asymmetry. This effect is very small until densities of the 
order of $\rho_0^0/10$ and progressively increases in the density regime 
corresponding to the inner crust.

Globally we can see that these in-medium modification are not negligible 
and should be accounted for in a realistic equation of state, in addition 
to the excluded volume mechanism.
Due to the simple expression (\ref{surf2}), these corrections can be 
tabulated as a function of $(A,I,\rho_{gn},\rho_{gp})$ and straightforwardly 
introduced in the NSE calculations as a modification of the cluster energy 
functional with no extra computational cost.

In this paper we focus on the qualitative effect of the binding energy 
shift on the equilibrium properties of the mixture between clusters and 
unbound nucleons, and on the uncertainties linked to out incomplete 
knowledge of the effective interaction. 
For this reason we do not explore the effect of this correction term on the 
whole $(T,\rho,Y_p)$ space, but limit ourselves to the simpler case of zero 
temperature stellar matter in $\beta$ equilibrium. 

\subsection{Effect of the in-medium correction at $\beta-$equilibrium 
at zero temperature}

\begin{figure*}
\resizebox{0.99\textwidth}{!}{
\includegraphics{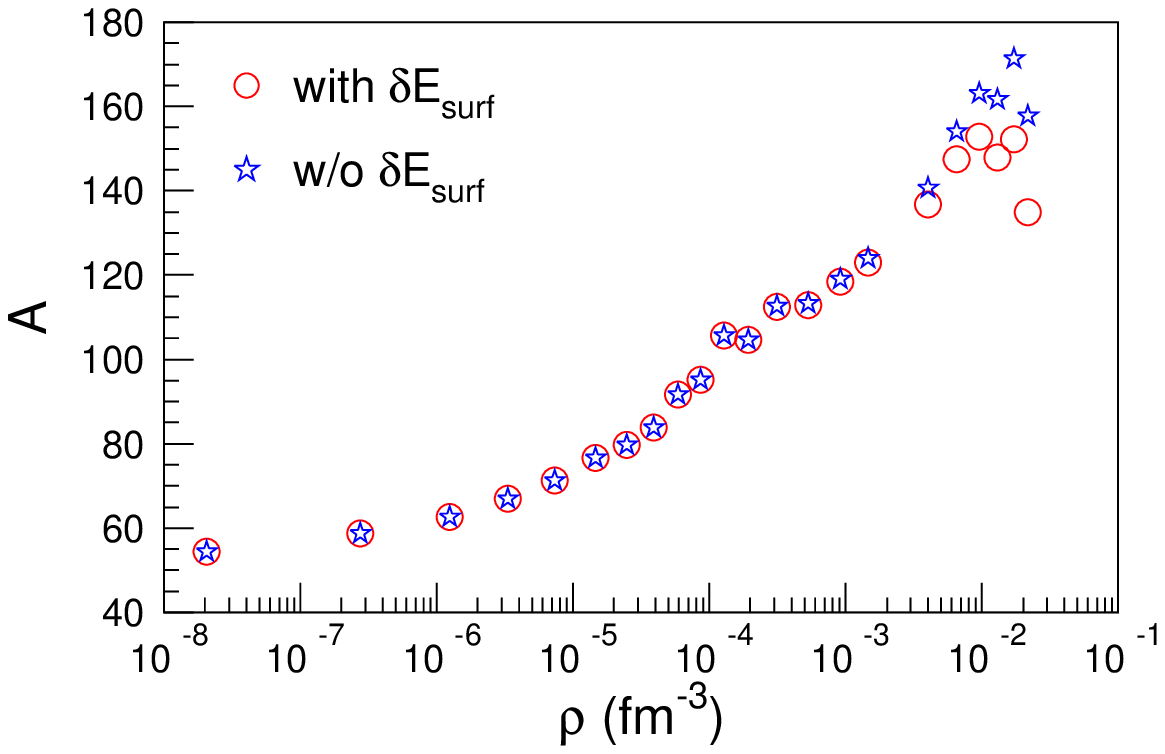}
\includegraphics{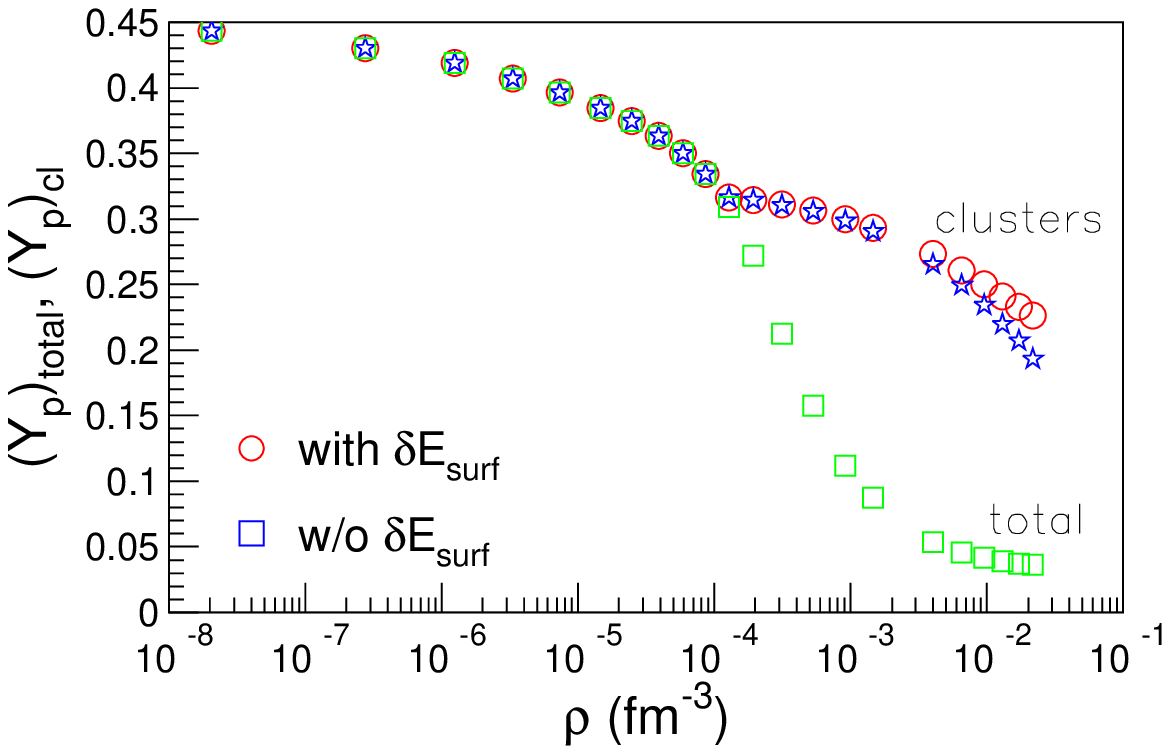}
}
\resizebox{0.99\textwidth}{!}{
\includegraphics{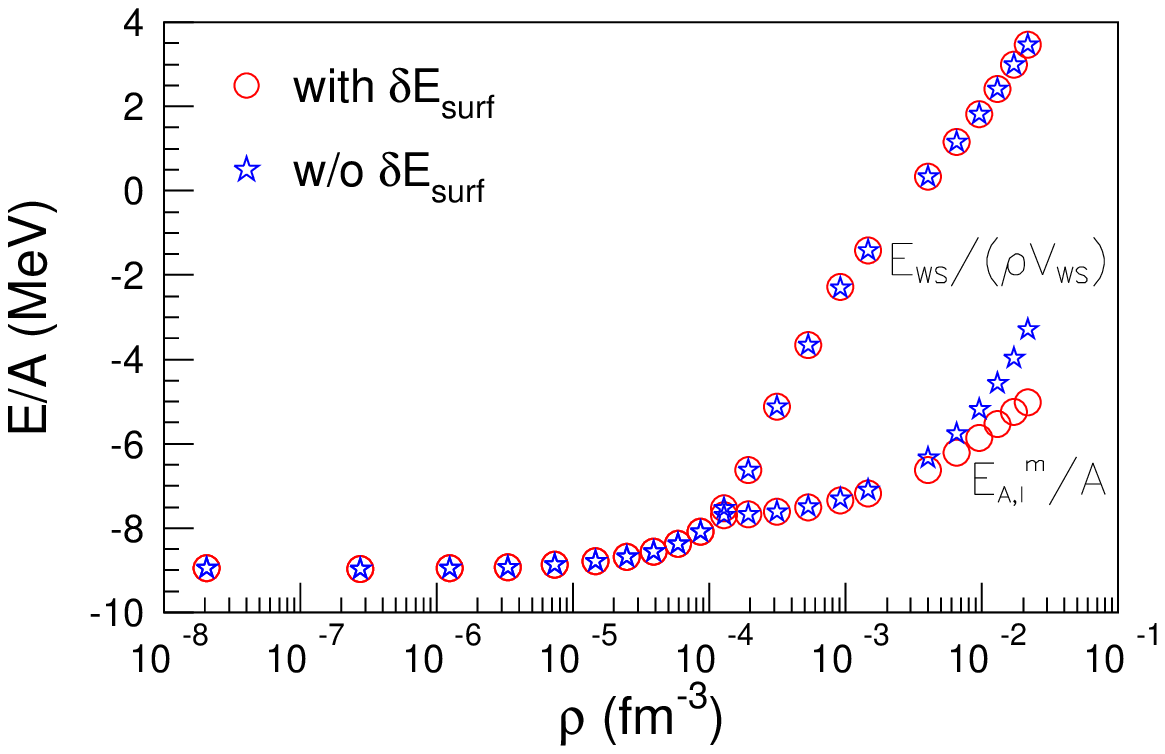}
\includegraphics{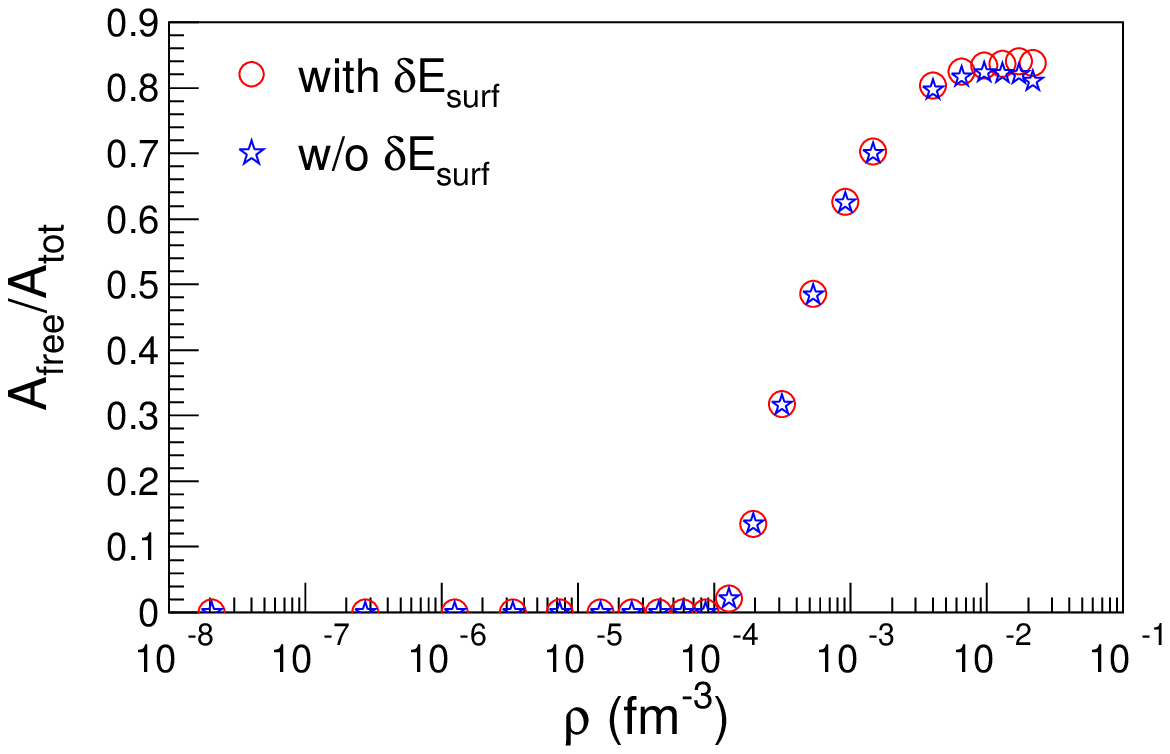}
}
\caption{Study of in-medium surface effects along the beta-equilibrium
path at $T$=0 MeV as a function of total baryonic density. 
Upper left: Cluster size; 
Upper right: Cluster proton fraction and total proton fraction; 
Lower left: Total energy per nucleon and cluster energy per nucleon;
Lower right: Percentage of unbound particles.
The considered effective interaction is SLY4.
}
\label{fig:Acl_T=0}
\end{figure*}

In order to estimate to which extent the crust-core transition of a neutron-star
and the related quantities are affected by in-medium effects, we have first 
calculated the Wigner-Seitz cells characteristics $(A_{WS},Z_{WS},V_{WS})$ 
at the different baryon densities as provided by NSE at 
$T$=0.5 MeV along the $\beta$-equilibrium path without 
the inclusion of the surface correction terms. We have verified 
that these numbers do not change by further decreasing the temperature 
and can thus be considered as representative of the zero temperature 
situation. The corresponding proton fraction as a function of the 
baryonic density is represented in the upper right part of 
Fig. \ref{fig:Acl_T=0}.

In this simplified situation  we can safely consider a pure neutron gas and
for each $\rho$, the Wigner-Seitz energy eq. (\ref{medium}) is minimized
with respect to the cluster size $A$ using the in-medium surface correction 
given by eq. (\ref{surf2}).
 
The equilibrium properties at zero temperature are plotted in 
Fig. \ref{fig:Acl_T=0}, both for the case where in-medium effects
are accounted for, and the one in which they are ignored.
We can see that the specificity of zero temperature is the discontinuous 
behavior of the number of unbound nucleons (lower right), 
which is strictly zero beyond the neutron drip-line and monotonically 
increases afterwards. As a consequence, both the cluster size (upper left)
and the cluster proton fraction (upper right) present a cusp behavior 
at the drip, allowing to clearly distinguish the inner crust from the 
outer crust. As we have seen in the previous section, this clear 
distinction is not possible at finite temperature, 
because of the presence of continuum states in the whole density domain.
Otherwise, the behavior is very similar to the trends discussed in section
\ref{section:NSE-betaeq} at $T=1 MeV$. 
This confirms the well-known fact that, at these low temperatures, 
the cluster distribution is well represented by the unique cluster 
obtained by the minimization of the energy, and the entropy contribution 
represents a correction on the global trend determined by the energetics 
of the Wigner-Seitz cell. 

In what regards the in-medium effects, Fig. \ref{fig:Acl_T=0} indicates
that they are sizable only at the densities corresponding to the inner crust, 
$\rho >5 \cdot 10^{-3}$ fm$^{-3}$ and they act in the sense of reducing 
the cluster size (upper right). This can be understood from the fact that 
a reduced surface energy especially favors clusters with a high surface 
to bulk ratio, that is small clusters. 
Since the medium is solely composed of unbound neutrons, a reduced cluster 
size corresponds to an increased isospin asymmetry (upper right). The effect 
of the in-medium modification appears globally small.
This is due to the fact that, at the densities where free nucleons can be 
found, zero temperature matter in $\beta$ equilibrium corresponds to 
extremely neutron-rich clusters in a neutron gas. As it can be seen from 
Fig. \ref{fig:deltaEsurf} (upper left) this situation 
corresponds to the smallest binding energy shift, and the excluded volume 
accounts for most of the in-medium effects. 
This might also explain why classical calculations which 
completely ignore this effect \cite{watanabe_2009,horowitz_prc2004} 
are still successful 
in reproducing the global phenomenology of the inner crust.
We expect that in supernova conditions the effect will be more important. 
Work in this direction is in progress.

The lower left part of Fig. \ref{fig:Acl_T=0} shows the total and cluster 
energy in the Wigner-Seitz cell. Again, we can clearly see the separation 
between outer and inner crust at the emergence of neutron drip, that leads 
to two different regimes for the density dependence of the cluster energy.
Because of the attractive nature of the interface interaction, the cluster 
energy is reduced by the in-medium surface effects. However, in the inner 
crust the unbound component is dominant as it can be seen from the lower 
right part. This implies that the total energy in the Wigner-Seitz cell 
is not affected by the in-medium surface corrections.
For this reason, the conclusions we have drawn on the sensitivity of the 
equation of state of clusterized matter in the previous section, 
where these effects were not accounted for, are expected to hold in a 
more sophisticated calculation of the cluster energy functional.

\section{Symmetry energy}

\begin{figure}
\resizebox{0.95\columnwidth}{!}{
\includegraphics{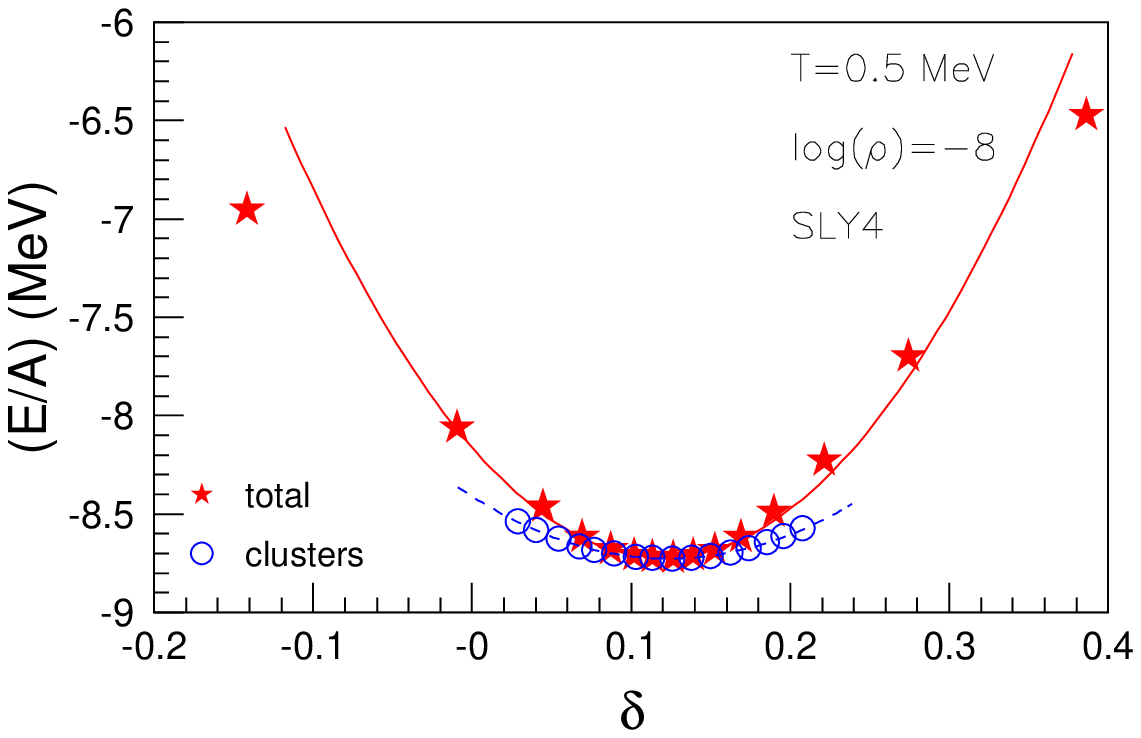}
}
\resizebox{0.95\columnwidth}{!}{
\includegraphics{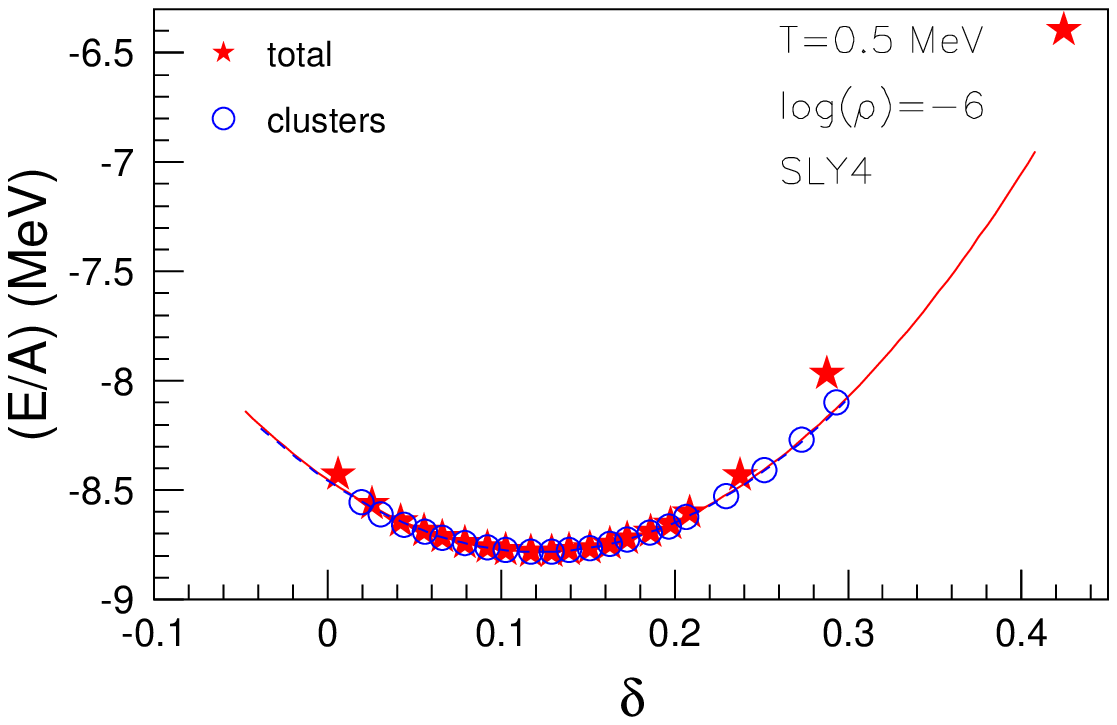}
}
\resizebox{0.95\columnwidth}{!}{
\includegraphics{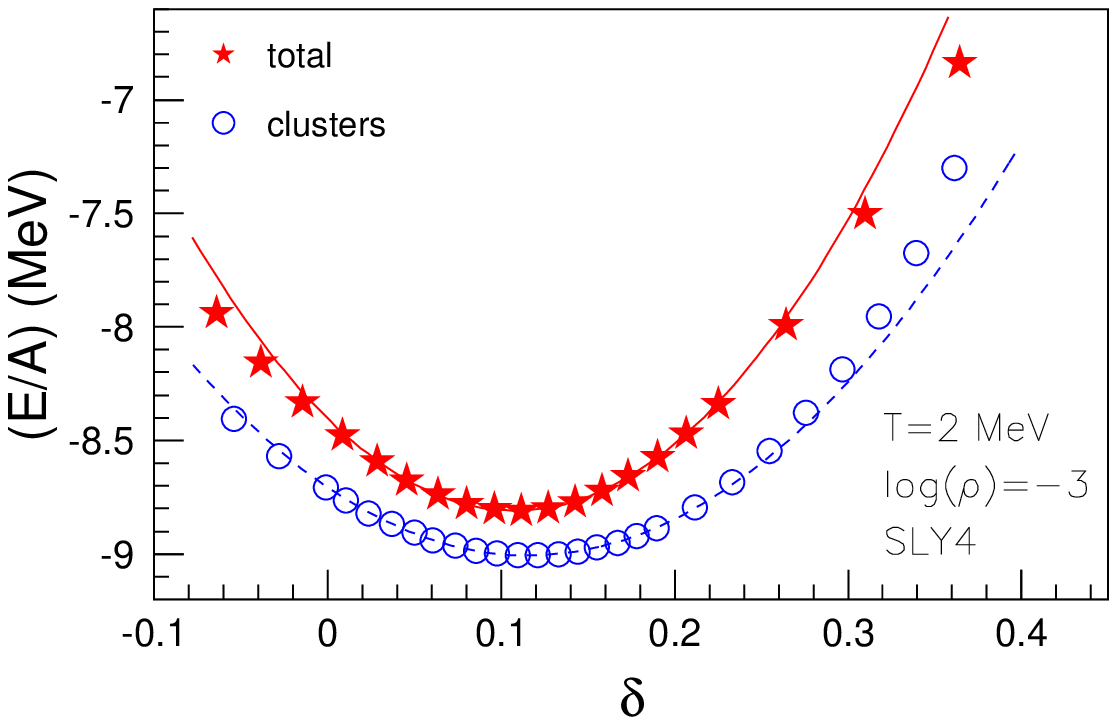}
}
\caption{Test of the quadratic approximation of the
energy per baryon versus asymmetry for different values of the 
total baryonic density and temperature. 
The total energy per baryon (stars) is plotted as well as the clusterized 
component value (open circles) as a function of the respective 
$\delta=1-2 Y_p$ values.
The employed effective interaction is SLY4.
The full and dashed lines correspond to a second order polynomial
fit of $e(\delta)$ and $e_{cl}(\delta_{cl})$, respectively.
}
\label{fig:quadratic_approx}
\end{figure}

One of the motivations of our study is the validity test,
in the case of dilute clusterized baryonic matter, of the parabolic
approximation of the energy per baryon as a function of density,
on which the definition of the symmetry energy relies.
The reason for which we expect this approximation to be violated
is the inhomogeneous structure of the matter, as we now explain.

%
%
%
%

The energy per baryon is a linear combination of the unbound 
$e_g=\epsilon/\rho_g$
and bound $e_{cl}=\langle (E_{A,I} + E^*_{A,I})/A \rangle_{\beta}$
contributions:
\begin{eqnarray}
\left (\frac EA \right )_{tot} = e(\rho,\delta)=e_{cl}(\langle A_{cl}\rangle_\beta,\langle\delta_{cl}\rangle_\beta,\rho_g,\delta_g) x_{cl} \nonumber \\
+ e_g(\rho_g,\delta_g) \left ( 1-x_{cl} \right )
\label{linear_comb}
\end{eqnarray}
where the expression of the total asymmetry as a function of the asymmetry of 
clusters and free particles depends in a complex way on the density and 
temperature:
\begin{equation}
\delta=x_{cl}\langle\delta_{cl}\rangle_\beta+(1-x_{cl})\delta_g \label{deltamix}
\end{equation}
We remind that the behavior of the unbound fraction as a function of the 
density was shown at zero and finite temperature in 
Figs. \ref{fig:Acl_T=0} and \ref{fig:cluster_prop_T=1} above 
in the specific case of $\beta$-equilibrium.

Two limiting situations can be considered.
In the limit $x_{cl}\ll 1$, $\rho\to\rho_g,\delta\to\delta_g$ and the energetics
of an homogeneous system is recovered. If we limit ourselves to the second 
order in $\delta$ and $T$ and note the Fermi energy 
$\epsilon_F=\hbar^2/2m(3\pi^2\rho/2)^{2/3}$, 
we get the standard mean-field result
\begin{equation}
\lim_{x_{cl}\to 0}e(\rho,\delta)= e_0(\rho,T)+e_{sym}(\rho,T)\delta^2  \label{symgas}
\end{equation}
where the isoscalar and isovector component depend on the effective 
interaction employed:
\begin{eqnarray}
e_0&=&C_0\rho+C_3\rho^{\sigma+1} 
+ \left ( \frac 35 \epsilon_F +\frac{\pi^2}{4}\frac{T^2}{\epsilon_F}\right )
\left ( 1+C_{eff}\frac{2m_0\rho}{\hbar^2}\right ), \nonumber \\
\end{eqnarray}
\begin{eqnarray}
e_{sym}&=&D_0\rho+D_3\rho^{\sigma+1} 
+ \frac 13\left (  \epsilon_F -\frac{\pi^2}{12}\frac{T^2}{\epsilon_F}\right )
\left ( 1+C_{eff}\frac{2m_0\rho}{\hbar^2}\right ) \nonumber \\
&+&D_{eff}\frac{2m_0\rho}{\hbar^2}\left (  \epsilon_F +
\frac{\pi^2}{12}\frac{T^2}{\epsilon_F}\right ).
\end{eqnarray}
%
%

In the opposite limit $x_{cl}\to 1$, 
$\rho \to A_0/V = \langle A_{cl}\rangle_\beta/\langle V_{WS}\rangle_\beta$,
$\delta\to\langle\delta_{cl}\rangle_\beta$ and 
the unbound particle energy, as well as the in-medium modification to the 
cluster energy, can be neglected.
The average energy per baryon is then determined by the finite temperature 
cluster energetics in the vacuum, 
which contains the isospin symmetry breaking Coulomb term:
\begin{eqnarray}
\lim_{x_{cl}\to 1}e(\rho,\delta)&=& e_{cl}=
\langle \frac {E_{A,I}}{A}\rangle_\beta +
\frac32 T \frac1{\langle A_{cl} \rangle_{\beta}}
\nonumber \\
& =& 
\tilde{a}_v(\langle A_{cl}\rangle_\beta,\rho_e,T)
\nonumber \\
&+&\tilde{a}_{sym}(\langle A_{cl}\rangle_\beta,\rho_e) \left ( \delta-\delta_0(\langle A_{cl}\rangle_\beta,\delta,\rho_e)\right )^2 \nonumber \\
&+&\frac32 T \frac1{\langle A_{cl} \rangle_{\beta}}
\label{ecl}
\end{eqnarray}
where again the isoscalar and isovector component depend on the effective 
interaction through the different terms of the cluster functional:
\begin{eqnarray}
\tilde{a}_v(A,\rho_e,T)=a_v-a_s A^{-1/3} \nonumber \\
-\frac{a_c}{4}f_{WS}(A,I,\rho_e)A^{2/3}(1-\delta_0)^2 + 
\langle \frac{E_{A,I}^*}{A}\rangle_\beta \\
\delta_0(A,\rho_e)=\frac{a_c f_{WS} A^{2/3}}{4a_a +a_c f_{WS} A^{2/3}} \label{delta0}
 \\
\tilde{a}_{sym}(A,\rho_e)=-a_a-\frac{a_c}{4} f_{WS} A^{2/3} \label{asym_cl}
\end{eqnarray}
We can see that in this limit a parabolic behavior is to be expected, 
but with a shifted minimum at a positive asymmetry due to the Coulomb term. 
This is true if the dependence on $\delta$ of $f_{WS}$ is sufficiently weak, 
which we expect to be true far from the crust-core transition.
In the general case, the weighted sum of the two parabolic behaviors will 
not give a parabola because of the non-linear dependence of $\delta$ on 
the cluster and free particles asymmetry eq.(\ref{deltamix}).

Fig. \ref{fig:quadratic_approx} shows the evolution of the cluster and total 
energy per baryon as a function of the $\delta$-isospin asymmetry parameter 
corresponding to different representative densities and temperatures. 
Qualitatively similar results are obtained for all the effective interactions 
explored in this work. In all cases a clear minimum is observed around 
$\delta\approx 0.1$, showing the important isospin symmetry breaking. 
This means that the usual definition of the symmetry energy as the curvature 
of the energy of symmetric matter in the isospin direction is not meaningful 
in star matter, and should be replaced by:
\begin{equation}
e_{sym}^{(1)}=\frac 12 \frac{\partial^2 e}{\partial \delta^2}|_{\delta=\delta_0(\rho)}
\label{sym1}
\end{equation}
where $\delta_0$ is the isospin asymmetry which minimizes the energy per 
baryon, due to the competition between the Coulomb and asymmetry terms eq. 
(\ref{delta0}).

We can see that at  low temperature the energetics is always dominated by the 
cluster component, and the limit $x_{cl}\to 1$ is approximately reached at 
densities as low as $\rho\geq 10^{-6}$ fm$^{-3}$ for $T=0.5$ MeV.
Increasing the temperature, the unbound component becomes more important in 
absolute value but still the global trend is determined by the bound clusters.
 Notice that this behavior of $x_{cl}$ with density and temperature 
is very different from the one observed in Figs.\ref{fig:Acl_T=0} and 
\ref{fig:cluster_prop_T=1}.
This is due to the fact that those figures were done in $\beta$-equilibrium, 
that is with an isospin asymmetry rapidly increasing with the density. 
Here we are interested in moderate $\delta\approx\delta_0$ asymmetries, where
the cluster fraction is dominant except at the lowest densities.

The visual behavior of the energy curves of Fig.\ref{fig:quadratic_approx} 
for moderate asymmetries appears parabolic at all densities
and temperatures. Concerning the cluster energy component, this is confirmed 
by a second order polynomial fit. 
This behavior can be understood considering that
the only non-parabolic term 
in eq.(\ref{ecl}) comes from the electron screening effect, $f_{WS}(I)$,
which depends on the baryon density but has a very weak dependence on 
$\delta$ for moderate asymmetries. 
Concerning the total energy, a closer analysis reveals that in the density 
domain where the free particles contribution cannot be neglected 
the curvature  of the total energy extracted from a parabolic fit depends 
strongly on the interval used for the fit, showing the presence of higher 
order contributions. This is again in agreement with our expectations 
from eq. (\ref{linear_comb}). 
 
\begin{figure}
\resizebox{0.99\columnwidth}{!}{
\includegraphics{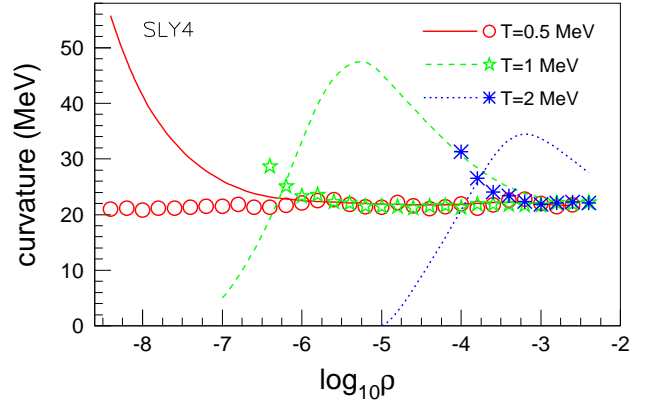}
}
\caption{Curvature of $(E/A)_{tot}$ (lines) and, respectively, 
$(E/A)_{cl}$ (symbols) in the 
direction of $\delta_{tot}$ and, respectively, $\delta_{cl}$ as a function of
total baryonic density at various temperatures, T=0.5, 1 and 2 MeV.
The considered effective interaction is SLY4.
}
\label{fig:curvature_tot_cl_T}
\end{figure} 

The global behavior of these symmetry energies as a function of density and 
temperature is displayed in Fig.\ref{fig:curvature_tot_cl_T}.
As we have seen in the previous chapter, at strictly zero temperature the free 
nucleons component appears only above the drip point. 
Because of the low density of the gas, this component gives a small energy 
contribution for all densities $\rho\leq 10^{-2}$ fm$^{-3}$. 
It is therefore not surprising that at low temperature the symmetry energy 
is dominated by the symmetry energy of the clusters. At variance with $T=0$, 
at finite temperature however free particles exist in equilibrium with clusters
at any density. As we have already observed in Fig. \ref{fig:quadratic_approx}, 
for moderate and constant asymmetries $x_{cl}$ is an increasing function of 
the density (with the exception of  the steep drop at the crust-core 
transition).
As a consequence, at the lowest densities the free particles component in 
eq. (\ref{linear_comb}) cannot be neglected.
This component is minimized at $\delta=0$, see eq.(\ref{symgas}). 
The presence of this shifted  behavior steepens the effective dependence on 
$\delta$ of the global system, leading to a higher symmetry energy with 
respect to the case of a nucleus in the vacuum
(see the solid line corresponding to T=0.5 MeV in Fig. 
\ref{fig:curvature_tot_cl_T}). 
At a given density, the importance of the unbound component increases with the 
temperature.
Above the solid-gas transition temperature, the opposite limit is recovered 
and the cluster component tends to disappear. 
We can see in Fig. \ref{fig:curvature_tot_cl_T} that this is the case 
at $T=1$ MeV for $\rho\leq 10^{-7}$ fm$^{-3}$ and
at $T=2$ MeV for $\rho\leq 2.5 \cdot 10^{-5}$ fm$^{-3}$
and homogeneous matter dominates the global energetics even at higher density.
As a consequence, isospin symmetry tends to be recovered at $T \geq 2$ MeV, 
the energy minimum is shifted towards $\delta=0$
and the symmetry energy essentially reflects the mean field behavior of a 
dilute gas. 
Finally, one can note that at the lowest densities at $T=1$ and 2 MeV 
the cluster symmetry energy deviates from the from the liquid-drop value. 
This stems from the kinetic energy term in eq. (\ref{ecl}) 
and, more precisely, the correlation among the average cluster size and
the total system asymmetry. 

\begin{figure}
\resizebox{0.99\columnwidth}{!}{
\includegraphics{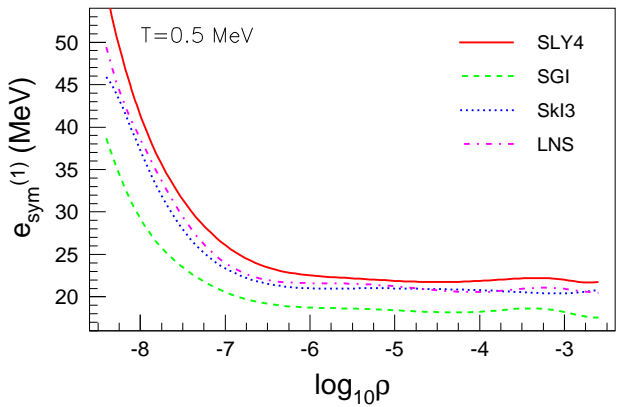}
}
\caption{$e_{sym}^{(1)}$
as a function of  baryonic density for T=0.5 MeV and different 
effective interactions.
{ The fitting interval is centered at $\delta_{0}$
and its width is $\Delta\delta=$0.1.} 
}
\label{fig:curvature_rho_T=1}
\end{figure}

 The sensitivity of the symmetry energy 
to the underlying effective interaction is plotted in Fig. 
\ref{fig:curvature_rho_T=1}.
The trends obtained employing SLY4 are confirmed by the other interactions.
While definitely exploring values far from both symmetry energy of saturated
symmetric matter and the one of most probable cluster, 
Eq. (\ref{sym1}) keeps the memory of the effective interaction. 
Indeed, the almost 3 MeV difference among the 
symmetry energy of saturated symmetric matter of SGI on one hand and
SLY4, SkI3 and LNS on the other hand is found in the difference the
various interactions provide for the symmetry energy of the inhomogeneous 
matter.
The different Skyrme models we have used represent roughly the present 
uncertainties on the isovector EoS properties. The difference between 
the different symmetry energies of inhomogeneous matter that we obtain 
thus gives a measure of how much this uncertainty in the effective 
interactions affects our knowledge of the symmetry energy properties 
of stellar matter.

\begin{figure}
\resizebox{0.99\columnwidth}{!}{
\includegraphics{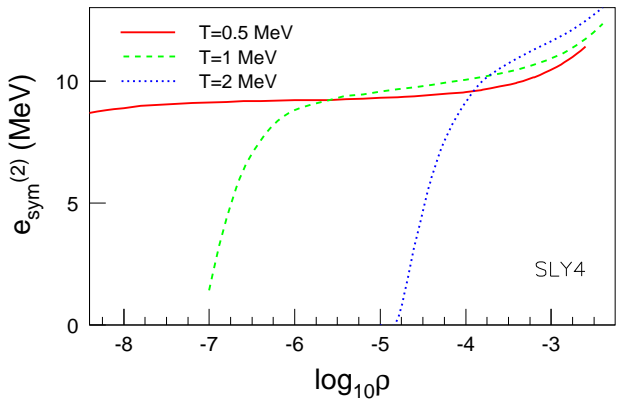}
}
\caption{The evolution with total baryonic density of $e_{sym}^{(2)}$
for T=0.5, 1 and 2 MeV.
The considered effective interaction is SLY4.
}
\label{fig:esym_a_la_typel}
\end{figure}

Another consequence of the isospin breaking Coulomb effect we have discussed,  
is that the definition of symmetry energy as an energy curvature by 
eq. (\ref{sym1}) will not be equivalent to the difference in binding 
between symmetric and neutron matter,
\begin{equation}
e_{sym}^{(2)}= e(\rho,\delta=1) - e(\rho,\delta=0)\neq
e_{sym}^{(1)}
\label{sym2}
\end{equation}
contrary to the common belief.  In particular, eq.(\ref{sym2}) was used to 
extract the symmetry energy 
of non-uniform matter in ref.\cite{typel}. The behavior of  eq.(\ref{sym2}) 
as a function of density and temperature is shown in 
Fig.\ref{fig:esym_a_la_typel}. Not surprisingly, this function has no 
ressemblance with the curvature at the energy minimum eq.(\ref{sym1}), 
and does not allow to infer the energy behavior of asymmetric clusterized 
matter, showing that these definitions should be handled with care.
Similar to ref.\cite{typel}, the presence of clusterization translates 
into a non-vanishing symmetry energy eq.(\ref{sym2})
in the $\rho\to 0$ limit.

\section{Conclusions}

In this paper we have analyzed the behavior of diluted stellar matter at 
zero and finite temperature in $\beta$-equilibrium in the framework of an 
improved NSE model.
The same effective interaction is consistently used to describe both unbound 
nucleons and nuclear clusters.
Bulk and surface in-medium modifications of the cluster energies are evaluated 
from the same effective interaction in the local density approximation. 
We have shown that the excluded volume effect exhausts the bulk part of the 
binding energy shift due to the presence of a medium. Surface corrections 
have a complex behavior as a function of the cluster size and isospin, 
and have to be consistently included in the NSE modelization in order to 
have a realistic equation of state. The net effect of this binding energy 
shift is to reduce the size of the clusters and modify the matter 
composition in the inner crust, while the global energetics is unmodified.

The presence of clusters at subsaturation densities leads to a deep 
modification of the global energetics, both in the isoscalar and in the 
isovector direction. Not only the baryonic energy is non-zero in the 
$\rho\to 0$ limit
\cite{typel}, but the parabolic approximation to the 
symmetry energy completely fails.
Indeed, the presence of charge fluctuations inside the globally charge-neutral 
medium induces important Coulomb effects which break the isospin invariance. 
As a consequence, the curvature of the energy functional in the isospin 
direction and the energy difference between neutron and symmetric matter 
diverge.

The other important effect of clusterization is that the effective density 
which is explored in stellar matter is different from the average baryonic 
density because of density fluctuations. As a consequence, the present 
uncertainties in the isovector part of the equation of state do not strongly 
affect the behavior of the equation of state of stellar matter, 
even if better constraints are certainly needed to have a 
fully quantitative prediction for astrophysical applications.
 
{\em Acknowledgements:}
This work has been partially funded by the SN2NS project 
ANR-10-BLAN-0503 and it has been supported by
Compstar, a research networking program of the European Science
foundation.
Ad. R. R acknowledges partial support from the Romanian National
Authority for Scientific Research under grant 
PN-II-ID-PCE-2011-3-0092 
and kind hospitality from LPC-Caen.

\end{document}